\documentclass[journal,letterpaper]{IEEEtran}
\usepackage{amsmath,amsfonts}
\usepackage{algorithmic}
\usepackage{algorithm}
\usepackage{array}
\usepackage[caption=false,font=normalsize,labelfont=sf,textfont=sf]{subfig}
\usepackage{booktabs}      % for 
\usepackage{wrapfig}
\usepackage{xcolor}         % colors
\usepackage{textcomp}
\usepackage{amssymb}
\usepackage{stfloats}
\usepackage{url}
\usepackage{verbatim}
\usepackage{graphicx}
\usepackage{subcaption}
\usepackage{mathtools}
\usepackage{cite}
\usepackage{pgfplots}
\usepackage{tabularx}
\usepackage[colorlinks=true, linkcolor=black, citecolor=black, urlcolor=blue]{hyperref}
\pgfplotsset{compat=1.16}
\usepackage{authblk} % This package allows for more complex author blocks

\begin{document}

\title{SHyPar: A Spectral Coarsening Approach to  Hypergraph Partitioning}
\author{Hamed Sajadinia$^{\star}$, Ali Aghdaei$^{\star}$, Zhuo Feng,~\IEEEmembership{Senior Member,~IEEE,}
        % <-this % stops a space
\thanks{$\star$ These authors contributed equally to this work.
\\The authors are with the Electrical and Computer Engineering Department, Stevens Institute of Technology, 1 Castle Point Terrace, Hoboken, NJ 07030, USA (e-mail: hsajadin@stevens.edu; aaghdaei@ucsd.edu; zfeng12@stevens.edu)}}

% \markboth{IEEE Transactions on Computer-Aided Design of Integrated Circuits and Systems, Vol. 43, N0. 10, OCTOBER 2024}%

% \IEEEpubid{0000--0000/00\$00.00~\copyright~2024 IEEE}


\maketitle

\begin{abstract}
State-of-the-art hypergraph partitioners utilize a multilevel paradigm to construct progressively coarser hypergraphs across multiple layers, guiding cut refinements at each level of the hierarchy. Traditionally, these partitioners employ heuristic methods for coarsening and do not consider the structural features of hypergraphs. In this work, we introduce a multilevel spectral framework, SHyPar, for partitioning large-scale hypergraphs by leveraging hyperedge effective resistances and flow-based community detection techniques. Inspired by the latest theoretical spectral clustering frameworks, such as HyperEF and HyperSF, SHyPar aims to decompose large hypergraphs into multiple subgraphs with few inter-partition hyperedges (cut size). A key component of SHyPar is a flow-based local clustering scheme for hypergraph coarsening, which incorporates a max-flow-based algorithm to produce clusters with substantially improved conductance. Additionally, SHyPar utilizes an effective resistance-based rating function for merging nodes that are strongly connected (coupled). Compared with existing state-of-the-art hypergraph partitioning methods, our extensive experimental results on real-world VLSI designs demonstrate that SHyPar can more effectively partition hypergraphs, achieving state-of-the-art solution quality.

\end{abstract}

\begin{IEEEkeywords}
Hypergraph Partitioning, Effective Resistance, Flow-Based Clustering, Spectral Coarsening.
\end{IEEEkeywords}

\section{Introduction}
\IEEEPARstart{H}{ypergraphs}
are more general than simple graphs since they allow modeling higher-order relationships among the entities. Hypergraph partitioning is an increasingly important problem with applications in various areas, including parallel sparse matrix computations \cite{catalyurek1999hypergraph}, computer-aided design (CAD) of integrated circuit systems \cite{karypis1999multilevel}, physical mapping of
chromosomes \cite{kucar2004hypergraph}, protein-to-protein interactions \cite{murgas2022hypergraph}, as well as data mining and machine learning \cite{ccatalyurek2022more}.

The problem of hypergraph partitioning involves grouping the nodes of a hypergraph into multiple clusters. This is done by minimizing a given cost function, defined over the hypergraph, under specific constraints. For instance, the cost function can be the hypergraph cut, which is the number (or total weight) of hyperedges that span more than one partition. The constraints may include balance constraints, where the difference in total node (or hyperedge) weight of each cluster should not exceed a given threshold. Due to these balance constraints, the problem of optimally partitioning a hypergraph is classified as NP-hard \cite{garey1979computers}.   

State-of-the-art multilevel hypergraph partitioning techniques predominantly rely on relatively simple heuristics for edge coarsening, such as vertex similarity or hyperedge similarity  \cite{karypis1999multilevel, devine2006parallel, vastenhouw2005two,ccatalyurek2011patoh,Shaydulin_2019}. For example, hyperedge similarity-based coarsening techniques contract similar large-size hyperedges into smaller ones. While this approach is straightforward to implement, it may adversely affect the original structural properties of the hypergraph. On the other hand, vertex-similarity-based algorithms determine strongly coupled (correlated) node clusters by measuring distances between vertices. This can be facilitated by hypergraph embedding, which maps each vertex to a low-dimensional vector, allowing for the Euclidean distance (coupling) between vertices to be easily computed in constant time. However, these simple metrics often fail to capture the higher-order global (structural) relationships within hypergraphs, potentially leading to suboptimal partitioning solutions.

Spectral methods are increasingly important in various graph and numerical applications \cite{teng2016scalable}. These applications include scientific computing and numerical optimization  \cite{spielman2014sdd,kelner2014almost,christiano2011flow},  graph partitioning and data clustering  \cite{spielmat1996spectral,lee2014multiway, peng2015partitioning}, data mining and machine learning \cite{kipf2016semi, defferrard2016convolutional}, graph visualization and data analytics \cite{koren2003spectral,Hu-CGA13,eades2017drawing},   graph signal processing and image segmentation \cite{shuman2013emerging,galasso2014spectral,ortega2018graph},  and integrated circuit (IC) modeling and verification \cite{xueqian:iccad14,lengfei:tcad15,zhuo:dac16,zhiqiang:dac17,zhiqiang:iccad17, zhuo:dac18}. Recent theoretical breakthroughs in spectral graph theory have led to the development of nearly-linear time spectral graph sparsification (edge reduction) \cite{spielman2011spectral,feng2020grass,Lee:2017,zhuo:dac18,kapralov2022spectral, kapralov2021towards, zhang2020sf} and coarsening (node reduction) algorithms \cite{loukas2018spectrally, zhao:dac19}.

On the other hand, spectral theory for hypergraphs has been less developed due to the more complicated structure of hypergraphs. For example, a mathematically rigorous approach has introduced a nonlinear diffusion process to define the hypergraph Laplacian operator, which measures the flow distribution within each hyperedge \cite{chan2018spectral,chan2020generalizing}; Additionally, the Cheeger's inequality has been validated for hypergraphs under this diffusion-based nonlinear Laplacian operator \cite{chan2018spectral}.  However, these theoretical results do not readily translate into practical implementations. Even recent breakthroughs, such as the SpecPart and K-SpecPart algorithms for spectral hypergraph partitioning \cite{bustany2023k,bustany2022specpart}, still depend heavily on the initial solutions provided by existing multilevel hypergraph partitioning methods, which themselves are based entirely on simplistic edge coarsening heuristics \cite{hmetis,kahypar}.

This paper introduces a brand-new multilevel hypergraph partitioning framework that leverages the latest highly scalable spectral hypergraph coarsening algorithms \cite{aghdaei2021hypersf, aghdaei2022hyperef}. Unlike traditional methods that depend solely on simple hyperedge (node) contraction heuristics and focus on local hypergraph structures, our framework, for the first time, incorporates spectral (global) properties into the multilevel coarsening and partitioning tasks as depicted in Figure \ref{fig:overview}. To achieve these goals, the paper is organized in a structured sequence of steps:

 \begin{figure*}[!t]
\centering
\includegraphics[width=0.789975\textwidth]{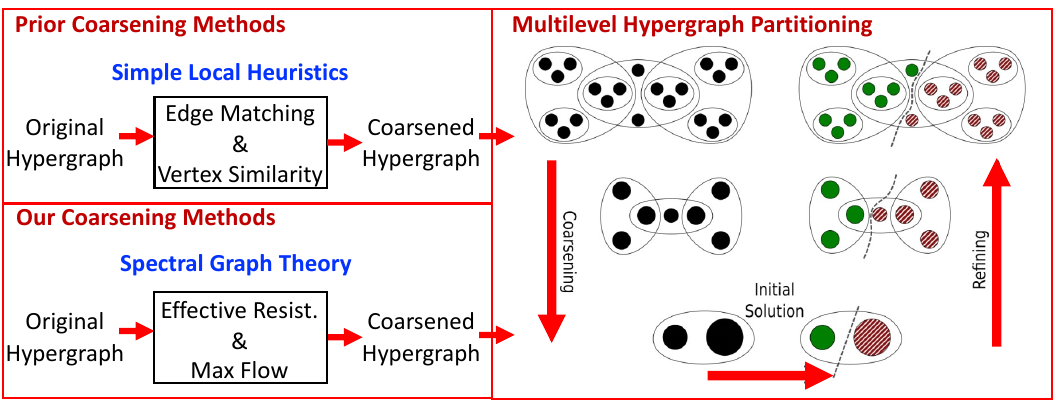}
\caption{The proposed multilevel hypergraph partitioning via spectral coarsening. }
\label{fig:overview}
\end{figure*}

 \paragraph{Scalable Spectral  Hypergraph Coarsening Algorithms}   This paper presents a two-phase scalable algorithmic framework for the spectral coarsening of large-scale hypergraphs, which exploits hyperedge effective resistances and strongly local flow-based methods \cite{aghdaei2021hypersf, aghdaei2022hyperef}. The proposed methods facilitate the decomposition of hypergraphs into multiple strongly-connected node clusters with minimal inter-cluster hyperedges, by incorporating the latest diffusion-based nonlinear quadratic operators defined on hypergraphs. 
 % \vspace{-10pt}

 \paragraph{Multilevel Hypergraph  Partitioning via Spectral Coarsening}  This paper develops a brand-new multilevel hypergraph partitioning tool by seamlessly integrating the proposed spectral coarsening methods into the hypergraph partitioning platform. By replacing the traditional simple coarsening heuristics with our theoretically rigorous spectral methods, the multilevel hypergraph partitioning tools developed through this research will potentially offer significantly improved partitioning solutions without compromising runtime efficiency.
  % \vspace{-10pt}
 
\paragraph{Validations of Multilevel  Hypergraph Partitioning Tools} This paper comprehensively validates the developed hypergraph partitioning tools, focusing on an increasingly important application: integrated circuit (IC) computer-aided design. Both the solution quality and runtime efficiency will be carefully assessed by testing the tools on a wide range of public-domain data sets. Additionally, the developed open-source software packages will be made available for public assessment.

The structure of this paper is organized as follows: Section II provides a foundational overview of the essential concepts and preliminaries in spectral hypergraph theory. Section III introduces the proposed method for hypergraph partitioning via spectral clustering, including resistance-based hypergraph clustering, flow-based clustering, and multilevel hypergraph partitioning. Section IV applies our framework to an established hypergraph partitioning tool, showcasing extensive experimental results on various real-world VLSI design benchmarks. The paper concludes with Section V, which summarizes the findings and implications of this work.

\section{Preliminaries and Background}

\subsection{Spectral (Hyper)graph Theory}
\subsubsection{Graph Laplacian matrix} In an undirected graph $G = (\mathcal{V}, \mathcal{E}, z)$, 
the symbol $\mathcal{V}$ represents a set of nodes (vertices), $\mathcal{E}$ represents a set of undirected edges, and $z$ indicates the weights associated with these edges. We denote ${D}$ as a diagonal matrix where each diagonal element ${D}(i,i)$ corresponds to the weighted degree of node $i$. Additionally, ${A}$ is defined as the adjacency matrix for the undirected graph $G$, as described below:

\begin{equation}\label{di_adjacency}
{A}(i,j)=\begin{cases}
z(i,j) & \text{ if } (i,j)\in \mathcal{E}  \\
0 & \text{otherwise }.
\end{cases}
\end{equation}
Subsequently, the Laplacian matrix of the graph $G$ is determined by the formula $L = D - A$. This matrix adheres to several key properties: (1) the sum of the elements in each row or column is zero; (2) all elements outside the main diagonal are non-positive; (3) the graph Laplacian is symmetric and diagonally dominant (SDD), characterized by non-negative eigenvalues.

\subsubsection{Courant-Fischer Minimax Theorem} The $k$-th largest eigenvalue of the Laplacian matrix $L \in \mathbb{R}^{|\mathcal{V}|\times|\mathcal{V}|}$ over the subspace $U$ of $\mathbb{R}^{\mathcal{V}}$, can be computed as follows:
\begin{equation}\label{eqn:minmax}
    \lambda_k(L) = \min_{dim(U)=k}\,{\max_{\substack{x \in U \\ x \neq 0}}{\frac{x^\top Lx}{x^\top x}}},
\end{equation}

This can be leveraged for compute the spectrum of the Laplacian matrix $L$.

 \subsubsection{Graph conductance} In a graph $G = (\mathcal{V}, \mathcal{E}, z)$ where vertices are partitioned into subsets $(S, \hat{S})$, the conductance of partition $S$ is defined as:
% \ali{conductance is definition, should not use equality symbol}
\begin{equation}\label{eqn:s_conductance}
    \Phi_G(S) \coloneqq \frac{w(S, \hat{S})}{\min\left(vol(S), vol(\hat{S})\right)} = \frac{\sum_{(i,j)\in \mathcal{E}:i\in S,j\notin S}{z(i, j)}}{\min\left(vol(S), vol(\hat{S})\right)},
\end{equation}
where the volume of the partition, $vol(S)$, is the sum of the weighted degrees of vertices in $S$, defined as $vol(S) := \sum_{i \in S}{d(i)}$. The graph's conductance \cite{chung1997spectral} is defined as:

\begin{equation}\label{eqn:G_conductance}
    \Phi_G := \min\limits_{\emptyset \not\subseteq S \subseteq \mathcal{V}} \Phi(S).
\end{equation}

\subsubsection{Cheegers' inequality} Research has demonstrated that the conductance $\Phi_G$ of the graph $G$ closely correlates with its spectral properties, as articulated by Cheeger's inequality \cite{chung1997spectral}:

\begin{equation}\label{eqn:cheeger}
    \omega_2/2 \leq \Phi_G \leq \sqrt{2\omega_2},
\end{equation}
where $\omega_2$ is the second smallest eigenvalue of the normalized Laplacian matrix $\widetilde{L}$, defined as $\widetilde{L} = D^{-1/2}LD^{-1/2}$.

\subsubsection{Effective resistance distance}Let $G = (\mathcal{V}, \mathcal{E}, z)$  represent a connected, undirected graph with weights $z \in \mathbb{R}^\mathcal{E}_{\geq 0}$. let  ${b_{p}} \in \mathbb{R}^{\mathcal{V}}$ denote the standard basis vector characterized by zero entries except for a one in the $p$-th position, and let    ${b_{pq}}=b_p-b_q$, The effective resistance between nodes $p$ and $q$, $(p, q) \in \mathcal{V}$ can be computed by: 
% \ali{The weights z should be associated with edges, not vertices. $z \in \mathbb{R}^\mathcal{E}_{\geq 0}$ is correct; Please check if we have made the same mistake in other parts.}
\begin{equation}\label{eq:eff_resist0}
    R_{eff}(p,q) = b_{pq}^\top L_G^{\dagger} b_{pq}=\sum\limits_{i= 2}^{|\mathcal{V}|} \frac{(u_i^\top b_{pq})^2}{\lambda_i}= \max\limits_{x \in \mathbb{R}^\mathcal{V}} \frac{(x^\top b_{pq})^2}{x^\top L_G x},
\end{equation}
where $L_G^{\dagger}$ represents the Moore-Penrose pseudo-inverse of the graph Laplacian matrix  $L_G$, and  $u_{i} \in \mathbb{R}^{\mathcal{V}}$ for $i=1,...,|\mathcal{V}|$ represents the  unit-length, mutually-orthogonal  eigenvectors corresponding to  Laplacian eigenvalues $\lambda_i$ for $i=1,...,|\mathcal{V}|$.

Graph conductance is a metric used to evaluate the quality of connectivity within a subset of nodes in a graph. Specifically, a lower conductance value indicates that the subset exhibits a higher number of internal edges relative to the number of edges connecting it to the remainder of the graph. This suggests that the subset forms a strongly-connected cluster with limited external interaction.
Alternatively, when a graph is modeled as a resistive electrical network, where nodes represent junctions and edge weights correspond to resistances, the concept of effective resistance between two nodes serves as a measure of their pairwise connectivity. A smaller effective resistance implies the presence of multiple alternative paths between the nodes, indicating stronger connectivity in the network.

\subsubsection{Spectral methods for hypergraphs}
% \textbf{Spectral Theory.} 
Classical spectral graph theory shows that the structure of a simple graph is closely related to the graph's spectral properties. Specifically, Cheeger's inequality demonstrates the close connection between expansion (or conductance) and the first few eigenvalues of graph Laplacians \cite{lee2014multiway}. Moreover, the Laplacian quadratic form computed with the Fiedler vector (the eigenvector corresponding to the smallest nonzero Laplacian eigenvalue) has been exploited to find the minimum boundary size or cut for graph partitioning tasks \cite{spielmat1996spectral}.
However, there has been very limited progress in developing spectral algorithms for hypergraphs. For instance, a classical spectral method has been proposed for hypergraphs by converting each hyperedge into undirected edges using star or clique expansions \cite{hagen1992new}. This naive hyperedge conversion scheme may result in lower performance due to ignoring the multi-way high-order relationships between the entities. A more rigorous approach by Tasuku and Yuichi \cite{soma2018spectral} generalized spectral graph sparsification for the hypergraph setting by sampling each hyperedge according to a probability determined based on the ratio of the hyperedge weight to the minimum degree of two vertices inside the hyperedge.
Another family of spectral methods for hypergraphs explicitly builds the Laplacian matrix to analyze the spectral properties of hypergraphs. A method has been proposed to create the Laplacian matrix of a hypergraph and generalize graph learning algorithms for hypergraph applications \cite{zhou2006learning}. A more mathematically rigorous approach by Chan et al. introduced a nonlinear diffusion process for defining the hypergraph Laplacian operator by measuring the flow distribution within each hyperedge \cite{chan2018spectral,chan2020generalizing}. Moreover, Cheeger's inequality has been proven for hypergraphs under the diffusion-based nonlinear Laplacian operator \cite{chan2018spectral}.

\subsubsection{Hypergraph  conductance} A hypergraph $H = (V, E, w)$  consists of a vertex set $V$ and a set of hyperedges $E$ with weights $w$. The degree of a vertex $d_v$ is defined as: $ d_v := \Sigma_{e \in E : v \in e} w(e)$, where $w(e)$ represents the weight of each hyperedge. The volume of a node set $S \subseteq V$ in the hypergraph is defined as: $vol(S) := \Sigma_{v \in S}d_v$. The conductance of a subset $S$ within the hypergraph is then calculated as:

\begin{equation}
    \Phi(S) := \frac{cut(S, \hat{S})}{min\{vol(S), vol(\hat{S})\}},
\end{equation}

where $cut(S, \hat{S})$ quantifies the number of hyperedges that cross between $S$ and $\hat{S}$. This computation uses an "all or nothing" splitting function that uniformly penalizes the splitting of hyperedges. The hypergraph's  overall conductance is defined as:
\begin{equation}
    \Phi_H := \min\limits_{\emptyset \not\subseteq S \subseteq V} \Phi(S),
\end{equation}

\subsection{Hypergraph Partitioning Methods}   
The previous hypergraph partitioners leverage a multilevel paradigm to construct a hierarchy of coarser hypergraphs using local clustering methods. Computing a sequence of coarser hypergraphs that preserve the structural properties of the original hypergraph is a key step in every partitioning method. The coarsening algorithm in existing partitioning methods either computes matching or clustering at each level by utilizing a rating function to cluster strongly correlated vertices. Hyperedge matching and vertex similarity methods are used in the coarsening phase to cluster the nodes and contract the hyperedges. Existing well-known hypergraph partitioners, such as hMETIS \cite{karypis1999multilevel}, KaHyPar \cite{kahypar}, PaToH \cite{ccatalyurek2011patoh}, and Zoltan \cite{devine2006parallel}, all use heuristic clustering methods to compute the sequence of coarser hypergraphs.

\subsubsection{Hypergraph coarsening}
 Multi-level coarsening techniques typically employ either matchings or clusterings on each level of the coarsening hierarchy. These algorithms utilize various rating functions to decide whether vertices should be matched or grouped together, using the contracted vertices to form the vertex set of the coarser hypergraph at the subsequent level. In contrast, n-level partitioning algorithms, such as the graph partitioner KaSPar \cite{osipov2010n}, establish a hierarchy of (nearly) n levels by contracting just one vertex pair between two levels. This approach eliminates the need for matching or clustering algorithms during the graph reduction process. KaSPar utilizes a priority queue to determine the next vertex pair to be contracted. After each contraction, it updates the priority of every neighboring vertex of the contracted vertex to maintain consistency in priorities. However, in hypergraphs, this method faces significant speed limitations because the size of the neighborhood can greatly expand due to a single large hyperedge. To address this limitation, KaHyPar \cite{kahypar} adopts the heavy-edge rating function. This strategy involves initially selecting a random vertex $p$ and contracting it with the best neighboring node that has the highest rating. The rating function specifically selects a vertex pair ($p,q$) that is involved in a large number of hyperedges with relatively small sizes, optimizing the coarsening process based on the most significant connections between vertices: 
 \begin{equation}
 \label{eq:rating_fuc}
    r(p,q) = \sum_{e=(p,q) \in E}{\frac{w(e)}{|e| - 1}},
\end{equation}
where $r(p,q)$ is the rating of the vertex pair, $w(e)$ is the weight of hyperedge $e$, and $|e|$ is the hyperedge cardinality.
\vspace{10 pt}
 \subsubsection{Community Detection}The coarsening phase aims to generate progressively smaller yet structurally consistent approximations of the input hypergraph. However, certain scenarios may arise where the inherent structure becomes obscured. For example, tie-breaking decisions may be necessary when multiple neighbors of a vertex share the same rating. Consequently, to improve coarsening schemes, existing partitioners like KaHyPar utilize a preprocessing step involving community detection to guide the coarsening phase. In this approach, the hypergraph is divided into several communities and then the coarsening phase is applied to each community separately. Existing community detection algorithms, such as the Louvain algorithm, partition hypergraph vertices into communities characterized by dense internal connections and sparse external ones. This method reformulates the problem into a task of modularity maximization in graphs.

\subsubsection{Partitioning objectives}
Hypergraph partitioning extends the concept of graph partitioning. Its objective is to distribute the vertex set into multiple disjoint subsets while minimizing a specified cut metric and adhering to certain imbalance constraints. The process of dividing into two subsets is known as \textit{bipartitioning}, whereas dividing into multiple subsets, typically referred to as \textit{$k$-way partitioning}, involves partitioning into $k$ parts.
More formally, consider a hypergraph $H=(V,E, w)$, where $k$ is a positive integer (with $k \geq 2$) and  $\epsilon$ is a positive real number (where $\epsilon \leq \frac{1}{k}$). The objective of $k$-way balanced hypergraph partitioning is to divide $V$ into $k$ disjoint subsets $S = \{V_0, V_1, \dots, V_{k-1}\}$ such that:
\vspace{5pt}
\begin{itemize}
\item{$(\frac{1}{k}- \epsilon) W \leq \sum_{v \in V_i}{w_v} \leq (\frac{1}{k}+\epsilon) W $, for $0 \leq i \leq k-1$ }
\vspace{5pt}
\item{$cutsize_H{(S)} = \sum_{\{e|e \not\subseteq V_i  \text{ for  any }i\}}{w_e}$ is minimized} 
\end{itemize}
\vspace{5pt}

Here, $k$ represents the number of partitions, $W$ is the hypergraph total weight ($W = \sum_{v \in V}{w_v}$), $\epsilon$ denotes the allowable imbalance among the partitions, and each $V_i$ is a block of the partition. We denote $S$ as an $\epsilon$-balanced partitioning solution.

\section{SHyPar: Hypergraph Partitioning via Spectral Coarsening}
To address the limitations of existing hypergraph coarsening methods that rely on simple heuristics, we propose a theoretically sound and practically efficient framework for hypergraph coarsening. Specifically, we propose a two-phase spectral hypergraph coarsening scheme based on the recent research on spectral hypergraph clustering \cite{aghdaei2021hypersf, aghdaei2022hyperef}. \textbf{Phase A} utilizes spectral hypergraph coarsening (HyperEF) to decompose a given hypergraph into smaller node partitions with bounded effective-resistance diameters \cite{aghdaei2022hyperef}. This is followed by \textbf{Phase B}, which guides the coarsening stage using a flow-based community detection method (HyperSF) aimed at minimizing the ratio cut \cite{aghdaei2021hypersf}. Next,  we exploit the proposed two-phase spectral hypergraph coarsening method for multilevel hypergraph partitioning: the prior heuristic hypergraph coarsening schemes will be replaced by the proposed spectral coarsening methods to create a hierarchy of coarser hypergraphs that can preserve the key structural properties of the original hypergraph.

\subsection{Resistance-Based Hypergraph Coarsening (Phase A)}
We leverage the effective resistance parameter to coarsen the hyperedges successively by contracting the hyperedges with small effective resistance. Effective resistance in simple graphs has been used to detect the critical edges for the global structure and it shows how different graph components are connected with each other.

\begin{figure}[ht!]
  \centering
  \includegraphics[width=0.6\linewidth]{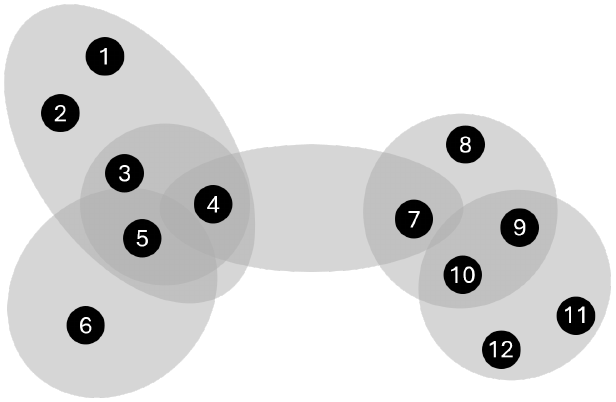}  
  \caption{The example of contraction in a simple hypergraph.}
  \label{fig:main-figure}
\end{figure}

\paragraph{Limitations of Existing Coarsening Methods}
The existing coarsening algorithms contract vertices at each level of the hierarchy. The primary method involves contracting each vertex with the best neighboring node. This is commonly done by using rating functions to identify and contract highly connected vertices.
However, these determinations often rely solely on the weights and sizes of the hyperedges. Rating functions, such as those described in Eq. (\ref{eq:rating_fuc}), based on hyperedge size are limited to the local structural properties of the hypergraph and do not account for its global structure. In contrast, the effective-resistance diameter provides a more comprehensive criterion, which considers the global structure.
To illustrate this limitation, consider a scenario where the hypergraph contains a bridge with few nodes (small size hyperedge), as illustrated in Figure \ref{fig:main-figure}. Algorithms that use hyperedge size tend to inappropriately contract bridge nodes (node 4 and node 7). This contraction can lead to the collapse of the overall structure of the hypergraph. Since the effective resistance of a bridge is high, algorithms based on effective resistance do not contract these nodes and preserve the integrity of the hypergraph structure.

In \cite{kapralov2022spectral}, the authors introduced a spectral algorithm based on effective resistance to sparsify hypergraphs. This method achieves nearly-linear-sized sparsifiers by sampling hyperedges according to their effective resistances \cite{kapralov2022spectral}. Despite its theoretical appeal, the technique involves a non-trivial procedure for the estimation of hyperedge effective resistances, which could hinder practical efficiency. The need to convert hypergraphs into graphs through clique expansion and the iterative updating of edge weights significantly adds to the complexity of the algorithm \cite{kapralov2022spectral}.

\begin{figure}
    \centering
    \includegraphics [width = 0.9\linewidth]{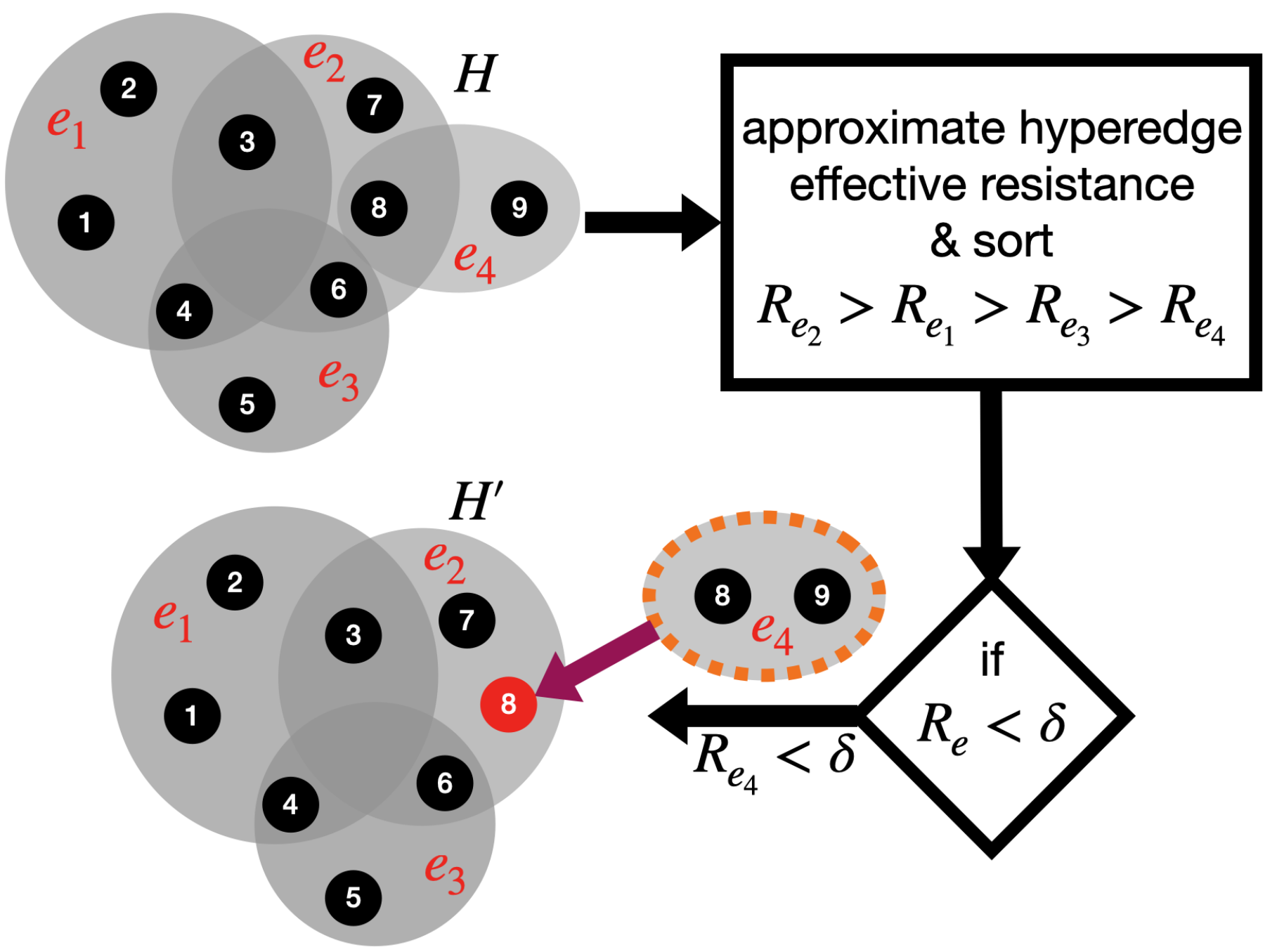}
    \caption{Overview of the HyperEF method.}
    \label{fig:EF_overview}
\end{figure}
\vspace{10 pt}

\subsubsection{HyperEF: Hypergraph Coarsening via effective resistance Clustering}

A spectral hypergraph coarsening technique, HyperEF, clusters nodes within each hyperedge when those nodes have a low effective-resistance diameter, as illustrated in Figure \ref{fig:EF_overview}. This approach significantly reduces the hypergraph size while preserving the original structural characteristics. A crucial part of HyperEF is an efficient procedure for estimating hyperedge effective resistances, which adapts the optimization-based method from Eq. (\ref{eq:eff_resist0}) to hypergraphs. Specifically, the effective resistance of a hyperedge is determined by finding an optimal vector $\chi^*$ via the following optimization:
\begin{equation} \label{eq:eff_resist_hypergraph}
\begin{split}
  R_e(\chi^*) =\max\limits_{\chi \in \mathbb{R}^\mathcal{V}} \frac{(\chi^{\top} b_{pq})^2}{Q_H(\chi)},  \quad p,q\in e
\end{split}
\end{equation}
where the original quadratic form $x^\top L_G x$ in Eq. (\ref{eq:eff_resist0}) is replaced by the nonlinear quadratic form $Q_H(\chi)$ \cite{chan2018spectral}:
\begin{equation}\label{eq:non-linerQ}
    Q_H(\chi) := \sum\limits_{e\in E} w_e \max\limits_{u, v \in e} (\chi_u - \chi_v)^2.
\end{equation}
As shown in Figure \ref{fig:EF_overview}, HyperEF generates a significantly smaller hypergraph $H' = (V', E', w')$ from the original hypergraph $H = (V, E, w)$ utilizing effective resistances of the hyperedge, achieving reductions in the number of vertices, edges and weights ($|V'| < |V|$, $|E'| < |E|$ and $|w'|<|w|$).

\vspace{10pt}

\subsubsection{Low-Resistance-Diameter Decomposition}
Consider a weighted undirected graph $G = (\mathcal{V}, \mathcal{E}, z)$ with weights $z \in \mathbb{R}^\mathcal{E}{> 0}$ and sufficiently large $\gamma > 1$. The effective-resistance diameter is defined as $\max\limits_{u,v \in \mathcal{V}} R_{eff}(u,v)$. Recent studies demonstrate that it is possible to partition a graph $G$ into multiple node clusters $G[\mathcal{V}_i]$ with low effective-resistance diameters by removing only a small fraction of edges \cite{alev2018graph}:
\begin{equation}\label{them:res}
    \max\limits_{u,v \in \mathcal{V}_i} R_{eff_{G[\mathcal{V}_i]}}(u,v) \lesssim \gamma^3 \frac{|\mathcal{V}|}{z(\mathcal{E})}.
\end{equation}
Additionally, let $\Phi_G$ represent the conductance of $G$. Cheeger's inequality provides a way to establish a relationship between the effective-resistance diameter of the graph and its conductance \cite{alev2018graph}:
\begin{equation}\label{them:res1}
     \max\limits_{u,v \in \mathcal{V}}R_{eff}(u,v) \lesssim \frac{1}{\Phi_G^2}.
\end{equation}
Leveraging recent spectral hypergraph theory \cite{chan2018spectral,chan2020generalizing}, HyperEF extends the above theorems to hypergraphs. Inequality (\ref{them:res}) implies that it is possible to decompose a hypergraph into multiple (hyperedge) clusters that have small effective-resistance diameters by removing only a few inter-cluster hyperedges, while (\ref{them:res1}) implies that contracting the hyperedges (node clusters) with small effective-resistance diameters will not significantly impact the original hypergraph conductance.
Based on these theoretical foundations, HyperEF consists of the following three main steps:
\begin{itemize}
\item Constructing the Krylov subspace to approximate the eigensubspace related to the original hypergraph;
\item Estimating the effective resistance of each hyperedge by applying the proposed optimization-based method;
\item Constructing the coarsened hypergraph by aggregating node clusters with low effective-resistance diameters.
\end{itemize}
\vspace{10pt}
\subsubsection{Fast Estimation of Hyperedge Effective Resistances}\label{Fast_Estimation}
To achieve high efficiency, the search for $\chi^*$ in Eq. (\ref{eq:eff_resist_hypergraph}) is restricted to an eigensubspace represented by a few select Laplacian eigenvectors from the simplified graph derived from the original hypergraph. Consider $G_b = (\mathcal{V}_b, \mathcal{E}_b, z_b)$ as the bipartite graph equivalent to the hypergraph $H = (V, E, w)$, where $|\mathcal{V}_b| = |V| + |E|$, $|\mathcal{E}_b| = \Sigma_{e \in E}|e|$, and $z_b$ denotes the scaled edge weights: $z(e, p) = \frac{w(e)}{d(e)}$.

According to Eq.(\ref{eq:eff_resist0}), hyperedge effective resistances can be approximated by identifying a set of orthogonal eigenvectors that maximize Eq. (\ref{eq:eff_resist_hypergraph}).
\begin{figure}[htbp]
 % \vspace*{-10pt}
    \centering
    % \subfloat[W/o reordering.]{
         \includegraphics [  width=0.9\linewidth]{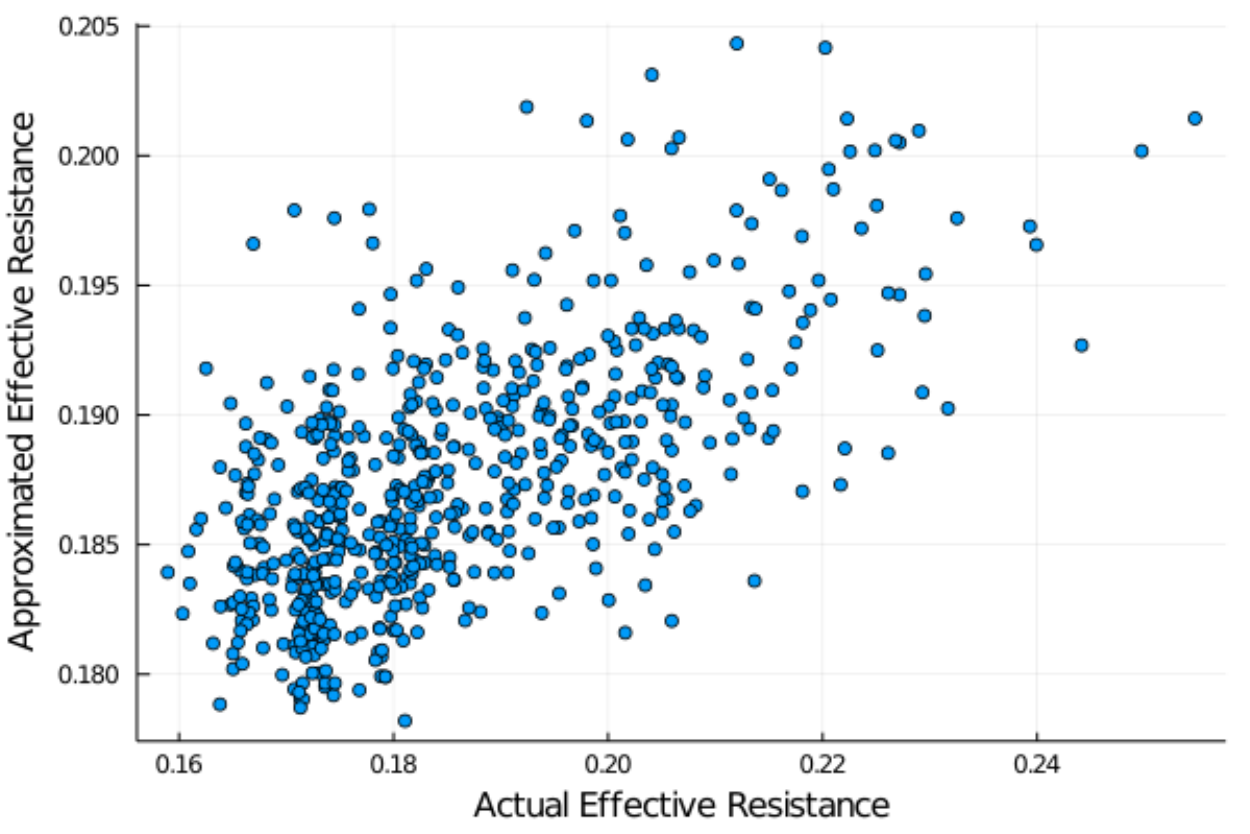}
    \caption{The approximated effective resistances for a simple graph.}
    \label{fig:cor_fig}
	% \vspace{-0pt}
\end{figure}
To avoid the computational complexity of determining eigenvalues and eigenvectors, a scalable algorithm is used in HyperEF to approximate the eigenvectors by leveraging the Krylov subspace, defined as follows:

For a non-singular matrix $A_{n \times n}$ and a non-zero vector $\chi \neq 0 \in \mathbb{R}^n$, the order-$(\rho+1)$ Krylov subspace generated by $A$ from $x$ is:
\begin{equation}\label{eq:krylov}
    \kappa_{\rho}(A, x) := span(x, Ax, A^2x, ..., A^{\rho}x),
\end{equation}
% \end{definition}
where $x$ is a random vector and $A$ is the normalized adjacency matrix of the simple graph obtained from the hypergraph via star expansion.

Let $x^{(1)}, x^{(2)}, ..., x^{(\rho)} \in \mathbb{R}^{\mathcal{V}_b}$ be the $\rho$ mutually-orthogonal vectors based on the order-$(\rho+1)$ Krylov subspace $\kappa_{\rho}(A, x)$. The effective resistance estimation in Eq.~(\ref{eq:eff_resist0}) is extended in HyperEF by incorporating the non-linear quadratic operator of hypergraphs from Eq.~(\ref{eq:non-linerQ}) to include the spectral properties of the hypergraph. By excluding the node embedding values associated with the star nodes in $x^{(i)}$, we generate a new set of vectors $\chi^{(i)}$ that are all mutually orthogonal. Each node in the hypergraph can then be embedded into a $\rho$-dimensional space. The resistance ratio ($r_e$) associated with a vector $\chi^{(i)} \in \chi^{(1)}, ..., \chi^{(\rho)}$ for each hyperedge $e$ is computed as:

\begin{equation}\label{eq:ratios}
     r_e(\chi^{(i)}) = \frac{(\chi^{(i)\top} b_{pq})^2}{Q_H(\chi^{(i)})},  \quad p,q\in e
\end{equation}
where $p$ and $q$ are the two maximally-separated nodes in the $\rho$-dimensional embedding space. Multiple resistance ratios $r_e^{(1)},..., r_e^{(\rho)}$ are returned by Eq. (\ref{eq:eff_resist_hypergraph}), corresponding to $\chi^{(1)},...,\chi^{(\rho)}$. After sorting resistance ratios in descending order, we have:
\begin{equation}
    r_e^1 > r_e^2 > ... > r_e^\rho.
\end{equation}
In HyperEF, the top $m$ resistance ratios are selected to estimate the effective resistance of each hyperedge. The hyperedge effective resistance ($R_e$) is specifically approximated by:
\begin{equation}\label{eq:R}
    R_e = \sum\limits_{i=1}^{m} r_e^i, \quad e \in E.
\end{equation}
Note that for each hypergraph, the Krylov subspace vectors in Eq. (\ref{eq:krylov}) only need to be computed once, which can be achieved in nearly linear time using only sparse matrix-vector operations. The effective resistance of each hyperedge can be estimated in constant time by identifying a few ($m$) Krylov subspace vectors that maximize the resistance ratio in Eq. (\ref{eq:ratios}). As shown in Figure \ref{fig:cor_fig}, the approximate effective resistances of simple graphs obtained using a few Krylov vectors correlate well with the ground-truth values, with a correlation coefficient of about 0.76.
Algorithm \ref{alg:effR} provides the detailed flow of the proposed hyperedge effective resistance estimation method.

\begin{algorithm}
\small { \caption{The effective resistance estimation algorithm flow}\label{alg:effR}}
\textbf{Input:} Hypergraph $H = (V,E, w)$, $\rho$.\\
\textbf{Output:} {A vector of effective resistance $R$ with the size $|E|$}.\\
  \algsetup{indent=1em, linenosize=\small} \algsetup{indent=1em}
    \begin{algorithmic}[1]
    \STATE Construct the bipartite graph $G_b$ corresponding to $H$. 
    \STATE Construct the order-$(\rho+1)$ Krylov subspace.
    \STATE Use Gram–Schmidt method to obtain the  orthogonal vectors.
    \STATE For each hyperedge compute its $\rho$ resistance ratios  using (\ref{eq:ratios}).
    \STATE Obtain all hyperedge effective resistances $R$ based on (\ref{eq:R}).
    % \STATE Orthogonalize all the vectors to each other.
     \STATE Return  $R$. 
    \end{algorithmic}
\end{algorithm}

\subsubsection{Hierarchical Effective Resistance Propagation }\label{Hierarchical}We preserve the hypergraph structure by employing a technique that iteratively computes a vector of effective resistance $R$ and contracts hyperedges exhibiting low effective resistances ($R_e < \delta$), where $\delta$ is the effective resistance threshold. Throughout this process, node clusters are contracted by merging the nodes within each cluster and replacing that cluster with a new $\textit{supernode}$ in the next level. By assigning a weight to each supernode—equal to the hyperedge's effective resistance evaluated at the previous level—we propagate essential structural information through all levels of coarsening. As an example, in Figure \ref{fig:EF_overview}, when hyperedge $e_4$ in $H$ is contracted, the node 8 in $H'$ takes on a weight that is equal to the effective resistance of $e_4$. This ensures that the structural information from the original hyperedge remains available in subsequent coarsening steps.

Let $H^{(l)} = (V^{(l)}, E^{(l)}, w^{(l)})$ represent the hypergraph at the $l$-th level. The vector of effective resistance $R$ is updated in each coarsening level according to the following equation:
\begin{equation}\label{eq:R_W_update}
     R_e^{(l)} \gets \sum_{v \in e} \eta(v)+R_e^{(l)},
\end{equation}
where $\eta(v)$ represents the weight of the nodes $v \in e$ corresponding to a contracted hyperedge from the previous level, initially set to all zeros for the original hypergraph. Consequently, the effective resistance of a hyperedge at a coarser level depends not only on $R_e^{(l)}$ (computed at the current level), but also on the effective-resistance data transferred from all previous levels. This allows the algorithm to preserve global structural information through the multilevel coarsening process.

The experimental results indicate that using Eq. (\ref{eq:R_W_update}) for effective resistances estimation yields more balanced hypergraph clustering outcomes compared to approaches that ignore the previous clustering information. The complete workflow for the effective resistance clustering algorithm, HyperEF, used for coarsening a hypergraph $H$ across $L$ levels, is detailed in Algorithm \ref{alg:HyperEF}.

\begin{algorithm}
\small { \caption{The HyperEF algorithm for hypergraph clustering  flow}\label{alg:HyperEF}}
\textbf{Input:} Hypergraph $H = (V,E, w)$, $\delta$, $L$, $\eta$.\\
\textbf{Output:} {A coarsened hypergraph $H' = (V', E', w')$ that $|V'| \ll |V|$}.\\
  \algsetup{indent=1em, linenosize=\small} \algsetup{indent=1em}
    \begin{algorithmic}[1]
    \STATE Initialize $H'$ $\leftarrow$ $H$
    \FOR{$l \gets 1$ to $L$}
    \STATE Call Algorithm \ref{alg:effR} to compute a vector of effective resistance $R$ with the size $|E'|$ for given hypergraph $H'$.
    \STATE Update the effective resistance vector $R$ by including the supernode weights using Eq. (\ref{eq:R_W_update})
    % \STATE Update the effective resistance vector $R$ by applying (\ref{eq:R_W_update}).
    \STATE Sort the hyperedges with ascending $R$ values.
    \STATE  Starting with the hyperedges that have the lowest effective resistances, contract (cluster) the hyperedge (nodes) if $R_e< \delta$.
    \STATE Construct a coarsened hypergraph $H'$ accordingly.
    \ENDFOR
    \STATE Return $H'$.
     
    \end{algorithmic}
\end{algorithm}

% \subsection{Phase B: Flow-based Hypergraph Cluster Refinement}
\subsection{Flow-Based  Community Detection (Phase B)}
To improve coarsening schemes for hypergraph partitioning, we utilize a community structure that integrates global hypergraph information into the coarsening process. This community structure directs the coarsening phase by permitting contractions solely within clusters.

In the flow-based community detection, we employ multilevel clustering through HyperEF, which enhances the spectral hypergraph clustering method by integrating a multilevel coarsening approach. Let $H = (V, E, w)$, the hypergraph  local conductance ($\textbf{HLC}$)  with respective to a   node-set $S$    is defined as follows \cite{veldt2020minimizing}:

\begin{equation}\label{obj:HLC}
    \textbf{HLC}_\mathcal{C}(S) = \frac{cut(S, \hat{S})}{vol(S \cap \mathcal{C}) - \beta vol(S \cap \hat{\mathcal{C}})},
\end{equation}

where $\mathcal{C} \subseteq V$ is reference node set, and $\beta$ is a locality parameter that modulates the penalty for incorporating nearby nodes outside set $\mathcal{C}$.

A spectral hypergraph coarsening algorithm (HyperSF) is proposed by minimizing the \textbf{HLC}, which has achieved promising results in hypergraph coarsening and partitioning in realistic VLSI designs.

\begin{figure*}[!t]
\centering
\includegraphics[width=0.89975\textwidth]{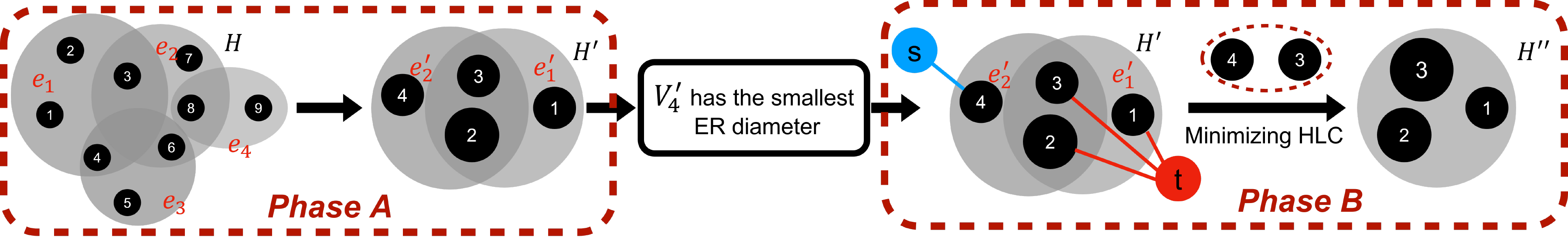}
\caption{Overview of the HyperSF method.}
\label{fig:hyperef}
\end{figure*}
\vspace{10pt}
 \subsubsection{Overview of   Coarsening Refinement (HyperSF)} 
\begin{figure}[htbp]
 % \vspace*{10pt}
    \centering
    % \subfloat[W/o reordering.]{
         \includegraphics [  width=0.568\linewidth]{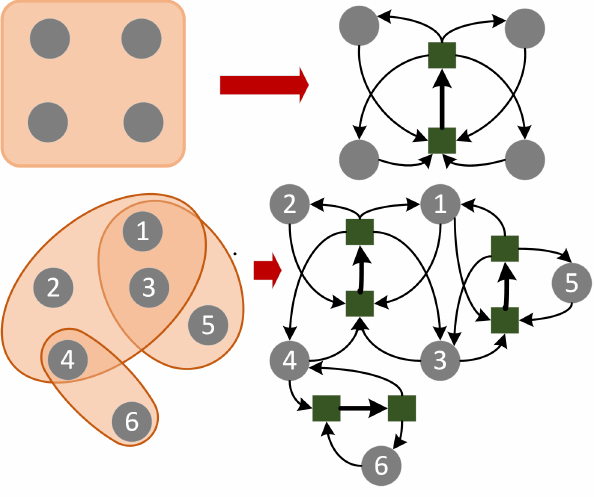}
    \caption{A hyperedge (top) and a hypergraph (bottom) converted to the corresponding directed graphs.}
    \label{fig:gadget}
	% \vspace{-20pt}
\end{figure}

Figure \ref{fig:hyperef} shows an overview of the HyperSF method. In this work, HyperSF is leveraged for only refining the most imbalanced node clusters (initially identified by HyperEF) with significantly smaller resistance diameters compared to the rest. Utilizing the flow-based clustering method that is proposed in \cite{veldt2020minimizing}, HyperSF aggregates strongly-coupled node clusters via minimizing Eq. (\ref{obj:HLC}): for each selected node set with large imbalance, HyperSF repeatedly solves a max $s$-$t$ flow, min $s$-$t$ cut problem to detect a set of neighboring node clusters that minimizes the local conductance \textbf{HLC} in Eq. (\ref{obj:HLC}). To this end,  the following key steps are applied (as shown in Figure \ref{fig:gadget}): \textbf{(Step 1)} an auxiliary hypergraph is constructed by introducing a source vertex $s$ and sink vertex $t$; \textbf{(Step 2)}  each hyperedge is replaced with a directed graph; \textbf{(Step 3)}  each seed node-set is iteratively updated by including new nodes into the set to minimize the \textbf{HLC} by repeatedly solving the  max $s$-$t$ flow, min $s$-$t$ cut problem: 
 \begin{equation}\label{eq:cutH}
    cut^{s-t}(S) = cut_H(S) + vol_H(\hat{S} \cap \mathcal{C}) + \beta vol_H(S \cap \hat{\mathcal{C}}).
\end{equation}
\textbf{(Step 4)} The node sets obtained from flow-based methods that minimize the local conductance are exploited to produce a smaller hypergraph with fewer nodes while preserving the key structural properties of the original hypergraph.  

 \begin{figure}[htbp]
 % \vspace*{-20pt}
    \centering
    % \subfloat[W/o reordering.]{
         \includegraphics [  width=0.6\linewidth]{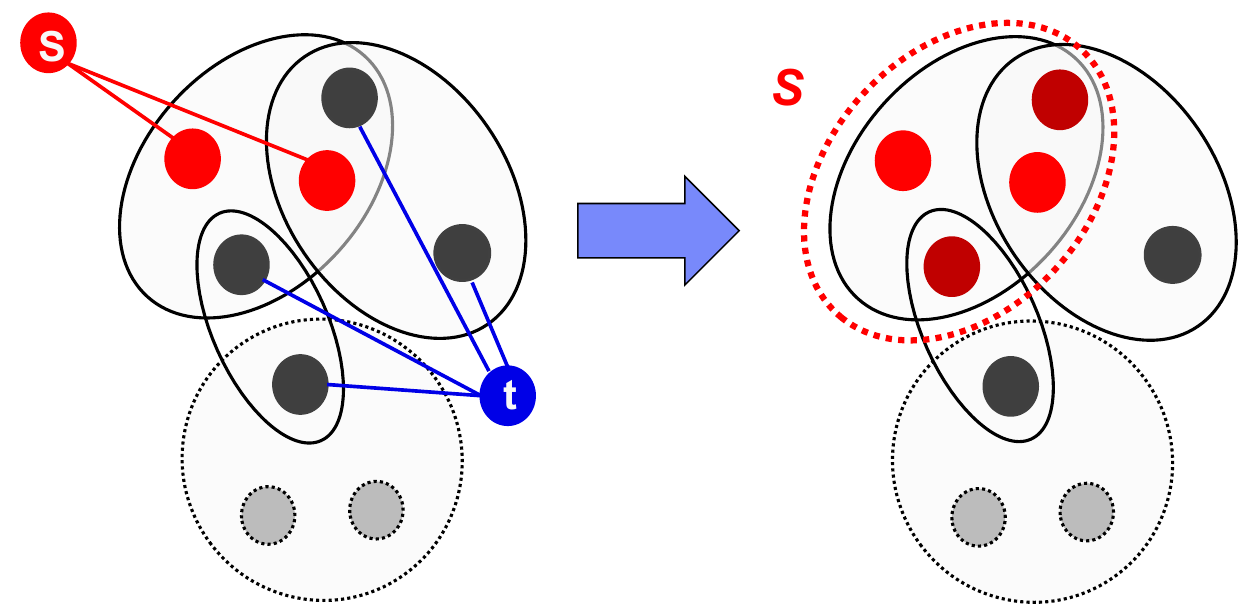}
    \caption{HyperSF minimizes \textbf{HLC} by iteratively solving  local max-flows.}
    \label{fig:hypersf}
	% \vspace{-0pt}
\end{figure}

 \vspace{10pt}

 \subsubsection{Local clustering algorithms}
The proposed flow-based clustering algorithm in \cite{veldt2020minimizing} is strongly local by expanding the network around the seed nodes $\mathcal{C}$, which benefits the coarsening framework in two ways: (1) applying the max $s$-$t$ flow, min $s$-$t$ cut problem on the local neighborhood of the seed nodes restricts node-aggregation locally and keeps the global hypergraph structure intact; (2) such a local clustering scheme significantly improves the algorithm efficiency due to the small-scale input dataset.  
\vspace{10pt}
 \subsubsection{Flow-based Local Clustering in HyperSF}
 First, we apply HyperEF (Algorithm \ref{alg:HyperEF}) to the hypergraph $H = (V,E, w)$ to compute a coarsened hypergraph $H' = (V',E', w')$ and identify isolated (unclustered) nodes, denoted by $\mathcal{C}$. These isolated nodes are simply those that remain unclustered after HyperEF’s multilevel coarsening step. Leveraging the introduced flow-based clustering in \cite{veldt2020minimizing}, HyperSF  constructs a sub-hypergraph $H'_L$ by iteratively expanding the hypergraph around the seed node set $\mathcal{C}$ and then repeatedly solve the hypergraph cut problem to minimize \textbf{HLC} until no significant changes in local conductance are observed. Define $E'(S) = \cup_{v' \in V', v' \in \{e'\}_{e' \in E'}}E'(v)$ for any set $S \subseteq V'$ and let $H'_\varsigma = (V' \cup \{s,t\}, E' \cup E'^{st})$, where $E'^{st}$ denotes the terminal edge set. HyperSF aims to construct a sub-hypergraph $H'_L$ to replace $H'_\varsigma$ that minimizes \textbf{HLC} by repeatedly solving a local version of (\ref{eq:cutH}). To this end, HyperSF sets up an oracle to discover a set of best neighborhood vertices for a given vertex $v'$:
\begin{equation}
    \kappa (v') = \{u' \in V'\}_{(u',v')\in e', e'\in E'}.
\end{equation}
HyperSF lets the oracle accept a set of seed nodes $\mathcal{C}$ and return $\kappa (\mathcal{C}) = \cup_{v' \in \mathcal{C}} \kappa (v')$. By utilizing the best neighborhood of the seed nodes $\kappa (\mathcal{C})$, HyperSF builds a local hypergraph $H'_L = (V'_L \cup \{s,t\}, E'_L \cup {E'}_L^{st} )$, where $V'_L = \mathcal{C} \cup \kappa (\mathcal{C})$ and $E'_L = \{e' \in E' \mid  V'_L \in e'\}$. As shown in Figure \ref{fig:hypersf}, HyperSF creates the local auxiliary hypergraph of $H'_L$ by introducing the source node $s$ and the sink node $t$, so that ${E'}_L^{st} \subseteq {E'}^{st}$ and repeatedly solves the max $s$-$t$ flow, min $s$-$t$ cut problem to minimize \textbf{HLC}.  The algorithm continuously expands   $H'_L$ and includes more vertices and hyperedges from $H'_\varsigma$ by solving (\ref{eq:cutH}) for the local hypergraph $H'_L$.

\begin{algorithm}
\small { \caption{Flow-based Hypergraph Clustering}\label{alg:flow_based}}
\textbf{Input:} Hypergraph $H = (V,E,w)$, and $\xi$ (Convergence parameter) \\
\textbf{Output:} {A set of vertices $S$ that minimizes \textbf{HLC(S)}};\\
    \algsetup{indent=1em, linenosize=\small} \algsetup{indent=1em}
    \begin{algorithmic}[1]
    \STATE Apply HyperEF to $H$ to compute coarsened hypergraph $H'= (V',E',w')$, and isolated nodes $\mathcal{C} \subseteq V$;
    \STATE Assign isolated nodes as seed nodes $S \gets \mathcal{C}$;
    \STATE $\Delta_{\textbf{HLC}} \gets \infty$;
    \WHILE{$\Delta_{\textbf{HLC}} > \xi$}
    \STATE Identify the best neighborhood of seed nodes $\kappa(S)$;
    \STATE Update $S$ according to $\kappa(S)$ to construct   $H'_L$; 
    \STATE Add a source node $s$ and sink node $t$ to $H'_L$;
    \STATE Repeatedly solve the max $s$-$t$ flow, min $s$-$t$ cut problem by minimizing (\ref{eq:cutH}) for $H'_L$;
    \STATE $\Delta_{\textbf{HLC}} \gets \textbf{HLC(S)}^j - \textbf{HLC(S)}^{j-1}$;
    \ENDWHILE
    \STATE Return $S$.
    \end{algorithmic}
\end{algorithm}

The algorithm \ref{alg:flow_based} presents the details of the flow-based local clustering technique. It accepts the original hypergraph $H=(V, E, w)$, and the convergence parameter $\xi$, which will output a set of strongly connected vertices $S$ that minimizes \textbf{HLC}.

 \subsection{Algorithm Complexity for Spectral  Coarsening}
 In HyperEF, the complexity for constructing the Krylov subspace for the bipartite graph $G_b = (\mathcal{V}_b, \mathcal{E}_b, z_b)$ corresponding to the original hypergraph $H= (V, E, w)$ is $O(|\mathcal{E}_b|)$; the complexity of the hyperedge effective resistance estimation and hyperedge clustering is $O(\rho|E|)$; the complexity of computing the node weights through the multilevel framework is $O(|E|)$ that leads to the overall nearly-linear algorithm complexity of $O(\rho|E|+|\mathcal{E}_b|)$. 
 In HyperSF, the runtime complexity of the proposed strongly local flow-based algorithm is $O\left(k^3 vol_H(\mathcal{C})^3(1 + \epsilon ^ {-1})^3\right)$, where $k$ is the maximum hyperedge cardinality. Since each phase in the proposed spectral hypergraph coarsening method has a nearly linear-time complexity, the entire two-phase spectral coarsening procedure will be highly scalable for handling large-scale hypergraphs.

\subsection{Hypergraph Partitioning with Spectral Coarsening and Flow-Based Clustering}
Building on previous hypergraph partitioning methods, we replace the coarsening algorithm with our proposed resistance-based approach and incorporate flow-based community detection to further enhance partitioning quality.

\subsubsection{Multilevel Spectral Coarsening with HyperEF}
While multilevel hypergraph partitioning frameworks like KaHyPar use hyperedge size metrics in their coarsening phase to determine which vertex pairs should be contracted, we introduce a resistance-based rating function that preserves structural properties when generating a hierarchy of smaller hypergraphs. For a given vertex pair $(p,q)$, we define the rating function as:
\begin{equation}\label{eq:ratingF}
r(p,q) = \sum_{\substack{e \in E \\ \{p, q\} \subseteq e}} \frac{w(e)}{R_e - 1}
\end{equation}
where $w(e)$ represents the weight of hyperedge $e$, and $R_e$ denotes its effective resistance. The effective resistance of each hyperedge, $R_e$ for $e \in E$, is computed using Eq.~(\ref{eq:R}) in nearly-linear time. The effective resistance information is further propagated throughout the multilevel coarsening scheme by using a node weight propagation technique, as described in  Eq. (\ref{eq:R_W_update}). This approach enables us to obtain a bound on the maximum distance between any pair of nodes within a hyperedge, thereby reducing the hypergraph size while maintaining the key spectral properties of the original hypergraph.\\

\subsubsection{Flow-Based clustering with HyperSF} To further improve the quality of partitioning, we incorporate flow-based community detection using HyperSF as described in Algorithm \ref{alg:flow_based}, as a preprocessing step. HyperSF creates high-quality clusters by analyzing the hypergraph flow structure.

The key insight of our approach is to first identify communities using HyperSF and then apply the multilevel spectral coarsening algorithm to each community separately. This community-aware coarsening strategy produces more balanced partitions that better preserve the hypergraph's structural properties.

HyperSF sorts clusters by their resistance diameter and prioritizes those with the lowest values. It then applies flow-based techniques to merge nearby clusters with low resistance diameters within the same neighborhood. The resulting clusters exhibit low hypergraph local conductance (\textbf{HLC}), which helps preserve the spectral properties of the original hypergraph during the coarsening process.

\subsubsection{SHyPar Algorithm Flow}
In Algorithm~\ref{alg:Shypar}, we present the detailed procedure of SHyPar, which leverages resistance-based spectral coarsening and flow-based clustering techniques for hypergraph partitioning.

The input is a weighted hypergraph $H = (V, E, w)$, and the output is a partitioned hypergraph where each node is assigned a partition index. \textbf{Line~1} represents the one-time computation of effective resistance using the HyperEF algorithm, which is employed during the coarsening phase. \textbf{Line~2} performs a one-time computation of flow-based clusters via the HyperSF algorithm, aiding in community detection. \textbf{Lines~3--5} outline the multilevel hypergraph partitioning pipeline, including coarsening, initial cut computation, and iterative refinement. \textbf{Line~6} produces the final partitioning solution of the hypergraph.

\begin{algorithm}
\small { \caption{SHyPar Algorithm}\label{alg:Shypar}}
\textbf{Input:} Hypergraph $H = (V,E,w)$ \\
\textbf{Output:} {Partitioned Hypergraph};\\
    \algsetup{indent=1em, linenosize=\small} \algsetup{indent=1em}
    \begin{algorithmic}[1]
    \STATE Setup: Hyperedge Effective resistance computation using HyperEF algorithm {\ref{alg:effR}}
    \STATE Setup: Flow-based community detection using HyperSF algorithm {\ref{alg:flow_based}}
    \STATE Multilevel spectral coarsening using Eq.{\ref{eq:ratingF}}
    \STATE Initial partitioning based on KaHyPar algorithm
    \STATE Solution refinement based on KaHyPar algorithm
    \STATE Return the partitioned hypergraph 
    \end{algorithmic}
\end{algorithm}

\section{Experimental Validation}
There are many applications related to hypergraph partitioning. This paper focuses on comprehensively evaluating the performance of the proposed hypergraph partitioning framework for increasingly important applications related to integrated circuits computer-aided design. Both the solution quality and runtime efficiency will be carefully assessed by testing them on a wide range of public-domain data sets. The implementation of the proposed algorithm, along with the code for reproducing the experimental results, is publicly available at \url{https://github.com/Feng-Research/SHyPar}.

To assess the performance of the proposed multilevel hypergraph partitioning tools in applications related to VLSI designs,  We apply the multilevel hypergraph partitioning tool developed to partition public-domain VLSI design benchmarks. For example, the  ISPD98 benchmarks that include ``IBM01'', ``IBM02'', ..., ``IBM18'' hypergraph models with   $13,000$ to $210,000$ nodes be adopted  \cite{alpert1998ispd98}.  The performance metrics for multilevel hypergraph partitioning, including the total hyperedge cut,  imbalance factors, and runtime efficiency, will be considered for comparisons with state-of-the-art hypergraph partitioning tools, such as hMETIS \cite{karypis1999multilevel}, and KaHyPar \cite{kahypar}.\\

\subsection{Spectral coarsening with HyperEF and HyperSF}
In this section, we compare the preliminary implementations of resistance-based hypergraph clustering (HyperEF) and flow-based clustering (HyperSF) with the well-known hypergraph partitioning tool, hMETIS \cite{karypis1999multilevel}.  Real-world VLSI design (hypergraph) benchmarks have been tested \cite{alpert1998ispd98}. All experiments have been evaluated on a computing platform with 8 GB of RAM and a 2.2 GHz Quad-Core Intel Core i7 processor.

\begin{figure}[htbp]
    \centering
         \includegraphics [  width=0.8\linewidth]{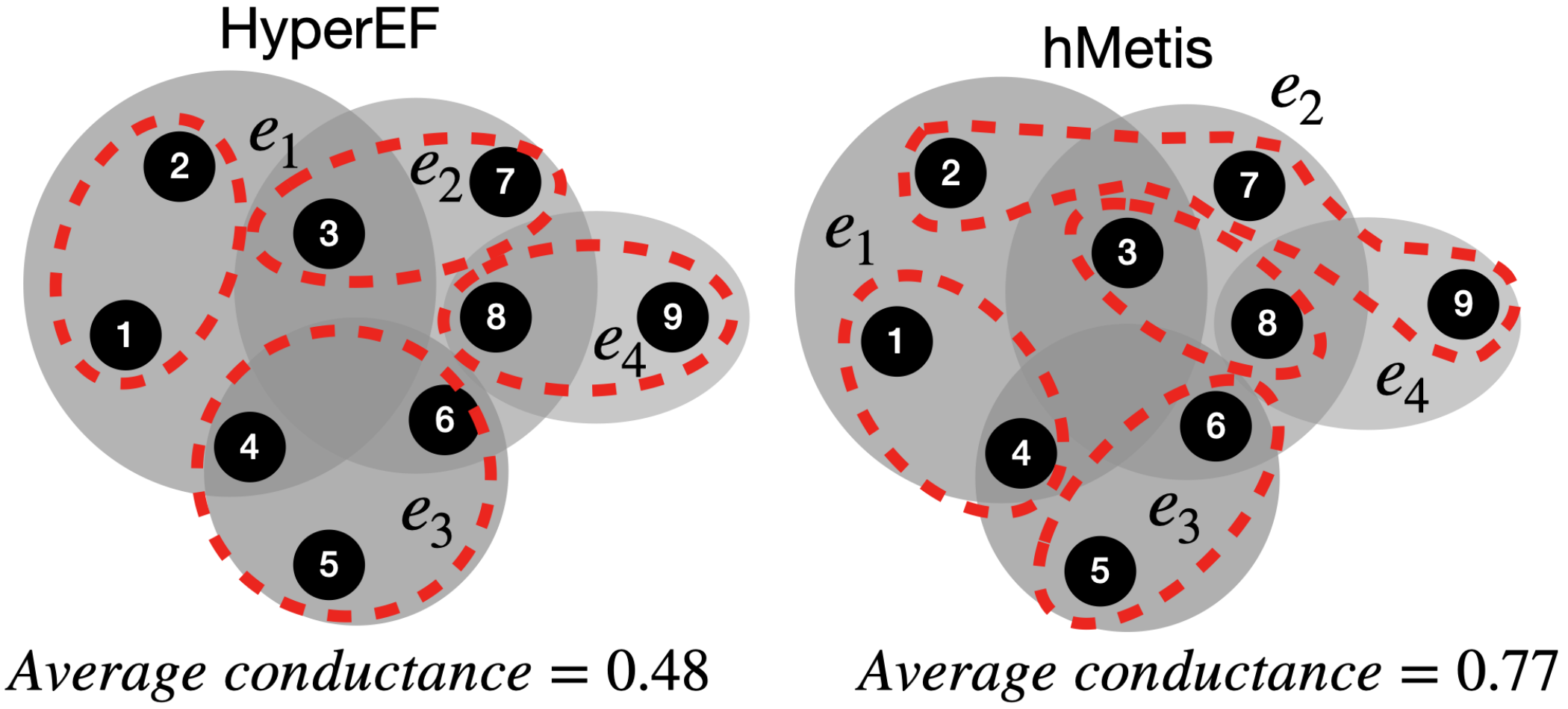}
    \caption{Resistance-based vs hMETIS Clustering Results. \protect\label{fig:hyperEFresult}}
\end{figure}

\subsubsection{HyperEF vs hMETIS for Hypergraph Coarsening}   HyperEF is compared to hMETIS for hypergraph coarsening considering both solution quality and runtime efficiency. The following average conductance of the node clusters is used to analyze the performance of each method. 
\begin{equation}
    \Phi_{\text{avg}} = \frac{1}{\left| S\right|}\sum\limits_{i=1}^{\left| S\right|} \Phi(S_i)
\end{equation}

Where $\Phi(S_i)$ denotes the   conductance of  node cluster $S_i$.  Figure \ref{fig:hyperEFresult} demonstrates the node clustering results for a small hypergraph obtained using HyperEF and hMETIS.  Both methods partition the hypergraph into four clusters, and the average conductance of node clusters has been computed to evaluate the performance of each method. The results show that HyperEF outperforms hMETIS by creating node clusters with a significantly lower average conductance. In addition, Table \ref{tab:hmetis}    shows the average conductance of node clusters $\Phi_{\text{avg}}$ computed with both HyperEF and hMETIS by decomposing the hypergraph   with the same node reduction ratios (NRs). With an NR = $75\%$ ($3\times$ node reduction)  HyperEF outperforms hMETIS in average conductance while achieving   $24-38\times$ speedups over hMETIS.

\begin{table}[ht]
% \tiny    
\caption{HyperEF vs hMETIS coductance (NR=75\%)} \label{tab:hmetis}
\centering
\footnotesize
\begin{tabular}{|c|c|c|c|c|}
\hline
Benchmark &
\begin{tabular}{@{}c@{}} $\Phi_{\text{avg}}$ \\
HyperEF
\end{tabular}      & \begin{tabular}{@{}c@{}} $\Phi_{\text{avg}}$ \\
hMETIS
\end{tabular}       &
\begin{tabular}{@{}c@{}} $\mathcal{T}$\scriptsize{(seconds)} \\
HyperEF
\end{tabular}  &
\begin{tabular}{@{}c@{}} $\mathcal{T}$\scriptsize{(seconds)} \\
hMETIS
\end{tabular}  \\ \hline
IBM01  & {\textbf{0.62}} & 0.65 & 1.23  & 29 (\textbf{24$\times$})     \\ \hline
IBM02  & {\textbf{0.62}} & 0.67 & 1.41  & 49 (\textbf{35$\times$})     \\ \hline
IBM03  & {\textbf{0.63}} & 0.66 & 2.11  & 53 (\textbf{25$\times$})     \\ \hline
IBM04  & {\textbf{0.64}} & 0.66 & 2.37  & 60 (\textbf{25$\times$})     \\ \hline
IBM05  & {\textbf{0.59}} & 0.63 & 2.34  & 62 (\textbf{26$\times$})     \\ \hline
IBM06  & {\textbf{0.64}} & 0.66 & 2.63  & 78 (\textbf{30$\times$})     \\ \hline
IBM07  & {\textbf{0.63}} & 0.67 & 3.54  & 115 (\textbf{32$\times$})    \\ \hline
IBM08  & {\textbf{0.61}} & 0.67 & 4.15  & 125 (\textbf{30$\times$})    \\ \hline
IBM09  & {\textbf{0.64}} & 0.66 & 4.38  & 131 (\textbf{30$\times$})    \\ \hline
IBM10  & {\textbf{0.63}} & 0.67 & 5.79  & 181 (\textbf{31$\times$})    \\ \hline
IBM11  & {\textbf{0.64}} & 0.67 & 5.73  & 176 (\textbf{31$\times$})    \\ \hline
IBM12  & {\textbf{0.65}} & 0.7  & 5.94  & 191 (\textbf{32$\times$})    \\ \hline
IBM13  & {\textbf{0.65}} & 0.68 & 6.87  & 229 (\textbf{33$\times$})    \\ \hline
IBM14  & {\textbf{0.62}} & 0.66 & 11.51 & 393 (\textbf{34$\times$})    \\ \hline
IBM15  & {\textbf{0.66}} & 0.69 & 14.44 & 486 (\textbf{34$\times$})    \\ \hline
IBM16  & {\textbf{0.63}} & 0.67 & 14.62 & 533 (\textbf{36$\times$})    \\ \hline
IBM17  & {\textbf{0.66}} & 0.7  & 15.22 & 568 (\textbf{37$\times$})    \\ \hline
IBM18  & {\textbf{0.6}}  & 0.67 & 15.79 & 602 (\textbf{38$\times$})    \\ \hline
\end{tabular}
\end{table}

\subsubsection{HyperSF vs hMETIS for Hypergraph Coarsening} In this section, we evaluate the performance of HyperSF against the hMETIS hypergraph partitioning tool. We measure the average local conductance $\text{\textbf{HLC}}_{\text{avg}}$ of the node clusters generated by each method, calculated as follows:
\begin{equation}
    \text{\textbf{HLC}}_{\text{avg}} = \frac{1}{\left| S\right|}\sum\limits_{i=1}^{\left| S\right|}\textbf{HLC}(S^i).
\end{equation}
Table \ref{tab:SF_hmetis} presents the average local conductance $\text{\textbf{HLC}}_{\text{avg}}$ for various methods under the same hypergraph reduction ratio (RR), where we reduce the number of nodes in each original hypergraph by $75\%$. The experimental data illustrate that HyperSF significantly enhances the average local conductance compared to the hMETIS method in all test scenarios.

\begin{table}[ht]
% \tiny    
\caption{HyperSF vs hMETIS local coductance (NR=75\%)} \label{tab:SF_hmetis}
\centering
\footnotesize
\begin{tabular}{|c|c|c|c|c|}
\hline
Benchmark &
\begin{tabular}{@{}c@{}} $\text{\textbf{HLC}}_{\text{avg}}$ \\
HyperSF
\end{tabular}      & \begin{tabular}{@{}c@{}} $\text{\textbf{HLC}}_{\text{avg}}$ \\
hMETIS
\end{tabular}       &
\begin{tabular}{@{}c@{}} $\mathcal{T}$\scriptsize{(seconds)} \\
HyperSF
\end{tabular}  &
\begin{tabular}{@{}c@{}} $\mathcal{T}$\scriptsize{(seconds)} \\
hMETIS
\end{tabular}  \\ \hline
IBM01  & {\textbf{0.44}} & 0.65 & 9.4 & 29   (\textbf{3$\times$})   \\ \hline
IBM02  & {\textbf{0.52}} & 0.69 & 22.6 & 49  (\textbf{2$\times$})   \\ \hline
IBM03  & {\textbf{0.48}} & 0.67 & 14.1 & 53  (\textbf{4$\times$})   \\ \hline
IBM04  & {\textbf{0.47}} & 0.68 & 15 & 60    (\textbf{4$\times$}) \\ \hline
IBM05  & {\textbf{0.55}} & 0.65 & 29.2 & 62  (\textbf{2$\times$})   \\ \hline
IBM06  & {\textbf{0.51}} & 0.68 & 30.1 & 78  (\textbf{3$\times$})   \\ \hline
IBM07  & {\textbf{0.48}} & 0.68 & 26.4 & 115 (\textbf{4$\times$})    \\ \hline
IBM08  & {\textbf{0.48}} & 0.68 & 43.3 & 125 (\textbf{3$\times$})    \\ \hline
IBM09  & {\textbf{0.47}} & 0.69 & 24.5 & 131 (\textbf{5$\times$})    \\ \hline
IBM10  & {\textbf{0.48}} & 0.68 & 45.2 & 181 (\textbf{4$\times$})    \\ \hline
IBM11  & {\textbf{0.46}} & 0.69 & 30.1 & 176 (\textbf{6$\times$})    \\ \hline
IBM12  & {\textbf{0.50}} & 0.71 & 42.4 & 191 (\textbf{5$\times$})    \\ \hline
IBM13  & {\textbf{0.48}} & 0.69 & 50.4 & 229 (\textbf{5$\times$})    \\ \hline
IBM14  & {\textbf{0.48}} & 0.67 & 85.7 & 393 (\textbf{5$\times$})    \\ \hline
IBM15  & {\textbf{0.47}} & 0.71 & 96 & 486   (\textbf{5$\times$})  \\ \hline
IBM16  & {\textbf{0.50}} & 0.70 & 116.8 & 533 (\textbf{5$\times$})     \\ \hline
IBM17  & {\textbf{0.51}} & 0.73 & 141.3 & 568 (\textbf{4$\times$})    \\ \hline
IBM18  & {\textbf{0.46}} & 0.68 & 129 & 602   (\textbf{5$\times$})  \\ \hline
\end{tabular}
\end{table}

\subsection{Hypergraph Partitioning with Spectral Coarsening} 
We compared SHyPar with leading hypergraph partitioners hMETIS \cite{karypis1999multilevel}, SpecPart \cite{bustany2022specpart}, KaHyPar \cite{kahypar}, and MedPart \cite{liang2024medpart} using two sets of publicly available benchmarks: the ISPD98 VLSI Circuit Benchmark Suite \cite{alpert1998ispd98} and the Titan23 Suite \cite{murray2013titan}. The details of these benchmarks are outlined in Table \ref{tab:table1} and Table \ref{tab:table_Titan23}. All tests were conducted on a server equipped with Intel(R) Xeon(R) Gold 6244 processors with 1546GB of memory.
% \vspace*{10pt}
\subsubsection{Experimental Setup}
To implement SHyPar, we have developed new hypergraph partitioning tools based on the existing open-source multilevel hypergraph partitioner, KaHyPar. We are utilizing the proposed two-phase spectral hypergraph coarsening method. Specifically, the heuristic coarsening scheme has been replaced by our novel spectral coarsening algorithm to create a hierarchy of coarser hypergraphs that preserve the key structural properties of the original hypergraph. Accordingly, we have substituted the existing coarsening method in KaHyPar with our proposed method, incorporating a new rating function. Additionally, the existing algorithm for community detection, Louvain, has been replaced with our proposed flow-based community detection method (Phase 2).

% \vspace*{10pt}
\subsubsection{SHyPar Performance on ISPD98 Benchmarks}
\begin{figure}[htbp]
    \centering
\begin{tikzpicture}
\begin{axis}[
    ylabel={Normalized Cutsize (\%)},
    symbolic x coords={IBM01, IBM02, IBM03, IBM04, IBM05, IBM06, IBM07, IBM08, IBM09, IBM10, IBM11, IBM12, IBM13, IBM14, IBM15, IBM16, IBM17, IBM18},
    xtick=data,
    x tick label style={rotate=90, anchor=east, font=\scriptsize}, % Labels are rotated and anchored at their eastern end
    ytick={70,75,...,110},
    ymin=70,
    ymax=112,
    legend pos=south west,
    ymajorgrids=true,
    xmajorgrids=true,
    grid style=dashed,
]

\addplot[
    color=red,
    mark=+,
    style=thick,
    mark options={solid},
    dashed
    ]
    coordinates {
    (IBM01,100)(IBM02,100)(IBM03,100)(IBM04,100)(IBM05,100)(IBM06,100)(IBM07,100)(IBM08,100)(IBM09,100)(IBM10,100)(IBM11,100)(IBM12,100)(IBM13,100)(IBM14,100)(IBM15,100)(IBM16,100)(IBM17,100)(IBM18,100)
};
\addlegendentry{KaHyPar}

\addplot[
    color=blue,
    mark=o,
    style=thick,
    dash pattern=on 10pt off 1pt
    ]
    coordinates {
    (IBM01,89.77)(IBM02,87.68)(IBM03,97.60)(IBM04,99.56)(IBM05,99.59)(IBM06,94.34)(IBM07,98.60)(IBM08,99.35)(IBM09,100.12)(IBM10,100.60)(IBM11,99.46)(IBM12,96.98)(IBM13,99.52)(IBM14,96.66)(IBM15,95.92)(IBM16,93.94)(IBM17,93.18)(IBM18,94.50)
};
\addlegendentry{SHyPar}
\addplot[
    color=black,
    mark=*,
    style=thick,
    dashdotted
    ]
    coordinates {
    (IBM01,100.74)(IBM02,97.43)(IBM03,101.05)(IBM04,105.21)(IBM05,104.23)(IBM06,111.44)(IBM07,101.67)(IBM08,106.32)(IBM09,106.34)(IBM10,106.03)(IBM11,103.79)(IBM12,98.82)(IBM13,106.18)(IBM14,108.57)(IBM15,98.66)(IBM16,110.41)(IBM17,96.77)(IBM18,98.37)
};
\addlegendentry{hMETIS}

\end{axis}
\end{tikzpicture}
\caption{ISPD98 benchmarks with unit weights $\epsilon$ = $1\%$ k = 4. \protect\label{fig:ISPDchart1}}
\end{figure}
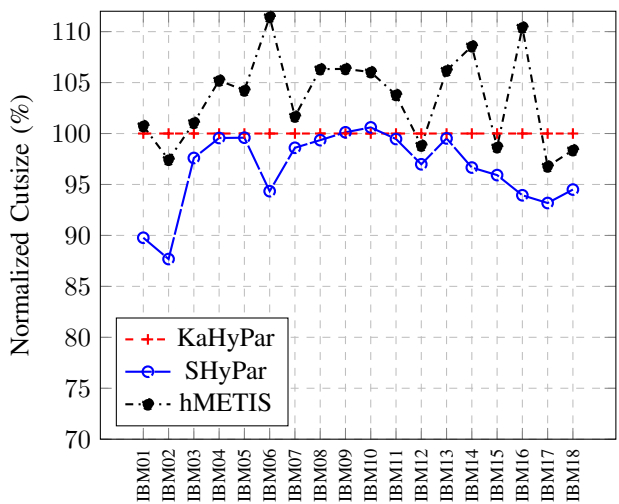

 Table \ref{tab:table1} presents a comparison of the cut sizes achieved by SHyPar on the ISPD98 VLSI circuit benchmark against those obtained from hMETIS, SpecPart, KaHyPar, and MedPart. The results for SHyPar show an average improvement of approximately 0.54\% for
 $\epsilon$ = $2\%$ and 0.4\% for $\epsilon$ = $10\%$, affirming its superiority over the best-published results. In several instances, SHyPar outperforms the best-published results by up to 5\%; these instances are specifically underlined for emphasis. Figure \ref{fig:ISPDchart} depicts the cut sizes obtained by SHyPar, KaHyPar, and hMETIS, normalized against the KaHyPar results. It is evident that SHyPar significantly enhances performance over both KaHyPar and hMETIS across many tests. Moreover, when SHyPar was applied to four partitions with $\epsilon$ = $1\%$, the improvements were consistent, as demonstrated in Figure \ref{fig:ISPDchart1}, which compares the cut sizes with those from KaHyPar and hMETIS, also normalized by KaHyPar results.

\begin{table*}[ht]
\caption{Statistics of ISPD98 VLSI circuit benchmark suite and cut sizes obtained by different approaches. The best results among all methods are highlighted in red.}
\label{tab:table1}
\centering
\footnotesize
\setlength{\tabcolsep}{4.46pt} % Reducing the cell padding horizontally
\begin{tabular}{|c|c|c||c|c|c|c|c||c|c|c|c|c|}
\hline
Benchmark  & \multicolumn{2}{|c|}{Statistics} & \multicolumn{5}{|c|}{$\epsilon$ = $2\%$} & \multicolumn{5}{|c|}{$\epsilon$ = $10\%$}\\
\hline
   &$|V|$&$|E|$& SpecPart & hMETIS & KaHyPar & MedPart & SHyPar & SpecPart & hMETIS & KaHyPar & MedPart & SHyPar\\
\hline
IBM01 & 12,752 & 14,111 & 202 & 213 & 202 & 202 & \textcolor{red}{\underline{201}} & 171 & 190 & 173 & \textcolor{red}{166} & \textcolor{red}{166}\\
\hline
IBM02 & 19,601 & 19,584 & 336 & 339 & 328 & 352 & \textcolor{red}{\underline{327}} & \textcolor{red}{262} & \textcolor{red}{262} & \textcolor{red}{262} & 264 & \textcolor{red}{262}\\
\hline
IBM03 & 23,136 & 27,401 & 959 & 972 & 958 & 955 & \textcolor{red}{\underline{952}} & 952 & 960 & \textcolor{red}{950} & 955 & \textcolor{red}{950}\\
\hline
IBM04 & 27,507 & 31,970 & 593 & 617 & \textcolor{red}{579} & 583 & \textcolor{red}{579} & \textcolor{red}{388} & \textcolor{red}{388} & \textcolor{red}{388} & 389 & \textcolor{red}{388}\\
\hline
IBM05 & 29,347 & 28,446 & 1720 & 1744 & 1712 & 1748 & \textcolor{red}{\underline{1707}} & 1688 & 1733 & \textcolor{red}{1645} & 1675 & \textcolor{red}{1645}\\
\hline
IBM06 & 32,498 & 34,826 & \textcolor{red}{963} & 1037 & \textcolor{red}{963} & 1000 & 969 & \textcolor{red}{733} & 760 & 735 & 788 & \textcolor{red}{733}\\
\hline
IBM07 & 45,926 & 48,117 & 935 & 975 & 894 & 913 & \textcolor{red}{\underline{882}} & \textcolor{red}{760} & 796 & \textcolor{red}{760} & 773 & \textcolor{red}{760}\\
\hline
IBM08 & 51,309 & 50,513 & 1146 & 1146 & 1157 & 1158 & \textcolor{red}{\underline{1140}} & 1140 & 1145 & \textcolor{red}{1120} & 1131 & \textcolor{red}{1120}\\
\hline
IBM09 & 53,395 & 60,902 & \textcolor{red}{620} & 637 & \textcolor{red}{620} & 625 & \textcolor{red}{620} & \textcolor{red}{519} & 535 & \textcolor{red}{519} & 520 & \textcolor{red}{519}\\
\hline
IBM10 & 69,429 & 75,196 & 1318 & 1313 & 1318 & 1327 & \textcolor{red}{\underline{1254}} &1261 & 1284 & 1250 & 1259 & \textcolor{red}{\underline{1244}}\\
\hline
IBM11 & 70,558 & 81,454 & 1062 & 1114 & 1062 & 1069 & \textcolor{red}{\underline{1051}} & 764 & 782 & 769 & 774 & \textcolor{red}{\underline{763}}\\
\hline
IBM12 & 71,076 & 77,240 & \textcolor{red}{1920} & 1982 & 2163 & 1955 & 1986 & 1842 & 1940 & \textcolor{red}{1841} & 1914 & \textcolor{red}{1841}\\
\hline
IBM13 & 84,199 & 99,666 & 848 & 871 & 848 & 850 & \textcolor{red}{\underline{831}} & 693 & 721 & 693 & 697 & \textcolor{red}{\underline{655}}\\
\hline
IBM14 & 147,605 & 152,772 & 1859 & 1967 & 1849 & 1876 & \textcolor{red}{\underline{1842}} & 1768 & 1665 & \textcolor{red}{1534} & 1639 & \textcolor{red}{1534}\\
\hline
IBM15 & 161,570 & 186,608 & 2741 & 2886 & 2737 & 2896 & \textcolor{red}{\underline{2728}} & 2235 & 2262 & \textcolor{red}{2135} & 2169 & \textcolor{red}{2135}\\
\hline
IBM16 & 183,484 & 190,048 & 1915 & 2095 & 1952 & 1972 & \textcolor{red}{\underline{1887}} & \textcolor{red}{1619} & 1708 & \textcolor{red}{1619} & 1645 & \textcolor{red}{1619}\\
\hline
IBM17 & 185,495 & 189,581 & 2354 & 2520 & \textcolor{red}{2284} & 2336 & 2285 & \textcolor{red}{1989} & 2300 & \textcolor{red}{1989} & 2024 & \textcolor{red}{1989}\\
\hline
IBM18 & 210,613 & 201,920 & 1535 & 1587 & 1915 & 1955 & \textcolor{red}{\underline{1521}} & 1537 & 1550 & 1915 & 1829 & \textcolor{red}{\underline{1520}}\\
\hline
\multicolumn{3}{|c|}{Average Improvement over hMETIS ($\%$)} & 3.64 & 0 & 2.19 & 1.11 & \textcolor{red}{4.92} & 2.91 & 0 & 2.77 & 1.65 & \textcolor{red}{4.77}\\
\hline
\end{tabular}
\end{table*}

\begin{figure}[htbp]
    \centering
\begin{tikzpicture}
\begin{axis}[
    ylabel={Normalized Cutsize (\%)},
    symbolic x coords={IBM01, IBM02, IBM03, IBM04, IBM05, IBM06, IBM07, IBM08, IBM09, IBM10, IBM11, IBM12, IBM13, IBM14, IBM15, IBM16, IBM17, IBM18},
    xtick=data,
    x tick label style={rotate=90, anchor=east, font=\scriptsize}, % Labels are rotated and anchored at their eastern end
    ytick={70,75,...,110},
    ymin=70,
    ymax=112,
    legend pos= south west,
    ymajorgrids=true,
    xmajorgrids=true,
    grid style=dashed,
]

\addplot[
    color=red,
    mark=+,
    style=thick,
    mark options={solid},
    dashed
    ]
    coordinates {
    (IBM01,100)(IBM02,100)(IBM03,100)(IBM04,100)(IBM05,100)(IBM06,100)(IBM07,100)(IBM08,100)(IBM09,100)(IBM10,100)(IBM11,100)(IBM12,100)(IBM13,100)(IBM14,100)(IBM15,100)(IBM16,100)(IBM17,100)(IBM18,100)
};
\addlegendentry{KaHyPar}

\addplot[
    color=blue,
    mark=o,
    style=thick,
    dash pattern=on 10pt off 1pt
    ]
    coordinates {
    (IBM01,99.5)(IBM02,99.7)(IBM03,99.37)(IBM04,100)(IBM05,99.7)(IBM06,100.6)(IBM07,98.65)(IBM08,98.53)(IBM09,100)(IBM10,95.14)(IBM11,98.96)(IBM12,91.81)(IBM13,97.99)(IBM14,99.62)(IBM15,99.67)(IBM16,96.67)(IBM17,100.04)(IBM18,79.42)
};
\addlegendentry{SHyPar}
\addplot[
    color=black,
    mark=*,
    style=thick,
    dashdotted
    ]
    coordinates {
    (IBM01,105.44)(IBM02,103.35)(IBM03,101.46)(IBM04,106.56)(IBM05,101.87)(IBM06,107.68)(IBM07,109.06)(IBM08,99.04)(IBM09,102.74)(IBM10,99.62)(IBM11,104.89)(IBM12,91.63)(IBM13,102.71)(IBM14,106.38)(IBM15,105.44)(IBM16,107.32)(IBM17,110.33)(IBM18,82.87)
};
\addlegendentry{hMETIS}

\end{axis}
\end{tikzpicture}
\caption{ISPD98 benchmarks with unit weights $\epsilon$ = $2\%$ k = 2. \protect\label{fig:ISPDchart}}
   % \label{fig:cor_fig}
	% \vspace{-20pt}
\end{figure}

\subsubsection{SHyPar Performance on Titan23 Benchmarks}

Building on the findings from the ISPD98 benchmarks, our research extended to the Titan23 benchmarks, notable for their high-degree hyperedges. Table \ref{tab:table_Titan23} details the results on the Titan23 benchmarks, where SHyPar dramatically outperforms hMETIS, achieving a 12\% improvement for $\epsilon$ = $2\%$ and an impressive 22.5\% for $\epsilon$ = $20\%$. Also SHyPar outperforms SpecPart, achieving a 3\% improvement for $\epsilon$ = $2\%$ and 1\% for $\epsilon$ = $20\%$. In cases such as sparcT2\_core, SHyPar even exceeds the best published results by up to 15\%; these notable achievements are highlighted by underlining.

\begin{table*}[ht]
\caption{Statistics of the Titan23 benchmark suite and cut sizes obtained by different approaches. The best results among all methods are highlighted in red.}
\label{tab:table_Titan23}
\centering
% \small
\footnotesize
\setlength{\tabcolsep}{4.46pt} % Reducing the cell padding horizontally
\begin{tabularx}{\textwidth}{|c|c|c||c|c|c|c|c||c|c|c|c|c|}
\hline
Benchmark  & \multicolumn{2}{|c|}{Statistics} & \multicolumn{5}{|c|}{$\epsilon$ = $2\%$} & \multicolumn{5}{|c|}{$\epsilon$ = $20\%$}\\
\hline
   &$|V|$&$|E|$& SpecPart & hMETIS & KaHyPar & MedPart & SHyPar & SpecPart & hMETIS & KaHyPar & MedPart & SHyPar\\
\hline
sparcT1\_core & 91,976 & 92,827 & 1012 & 1066 & \textcolor{red}{974} & 1067 & \textcolor{red}{974} & 903 & 1290 & 873 & \textcolor{red}{624} & 631\\
\hline
neuron & 92,290 & 125,305 & 252 & 260 & 244 & 262 & \textcolor{red}{\underline{243}} & \textcolor{red}{206} & 270 & 244 & 270 & 244\\
\hline
stereo\_vision & 94,050 & 127,085 & 180 & 180 & \textcolor{red}{169} & 176 & \textcolor{red}{169} & \textcolor{red}{91} & 143 & \textcolor{red}{91} & 93 & \textcolor{red}{91}\\
\hline
des90 & 111,221 & 139,557 & 402 & 402 & 380 & \textcolor{red}{372} & 379 & 358 & 441 & 380 & 349 & \textcolor{red}{\underline{345}}\\
\hline
SLAM\_spheric & 113,115 & 142,408 & \textcolor{red}{1061} & \textcolor{red}{1061} & \textcolor{red}{1061} & \textcolor{red}{1061} & \textcolor{red}{1061} & \textcolor{red}{1061} & \textcolor{red}{1061} & \textcolor{red}{1061} & \textcolor{red}{1061} & \textcolor{red}{1061}\\
\hline
cholesky\_mc & 113,250 & 144,948 & 285 & 285 & \textcolor{red}{283} & \textcolor{red}{283} & \textcolor{red}{283} & 345 & 667 & 591 & \textcolor{red}{281} & 479\\
\hline
segmemtation & 138,295 & 179,051 & 126 & 136 & \textcolor{red}{107} & 114 & \textcolor{red}{107} & \textcolor{red}{78} & 141 & \textcolor{red}{78} & \textcolor{red}{78} & \textcolor{red}{78}\\
\hline
bitonic\_mesh & 192,064 & 235,328 & \textcolor{red}{585} & 614 & 593 & 594 & 586 & \textcolor{red}{483} & 590 & 592 & 493 & 506\\
\hline
dart & 202,354 & 223,301 & 807 & 844 & 924 & 805 & \textcolor{red}{\underline{784}} & 540 & 603 & 594 & 549 & \textcolor{red}{\underline{539}}\\
\hline
openCV & 217,453 & 284,108 & 510 & 511 & 560 & 635 & \textcolor{red}{\underline{499}} & 518 & 554 & 501 & 554 & \textcolor{red}{\underline{473}}\\
\hline
stap\_qrd & 240,240 & 290,123 & 399 & 399 & \textcolor{red}{371} & 386 & \textcolor{red}{371} & 295 & 295 & \textcolor{red}{275} & 287 & \textcolor{red}{275}\\
\hline
minres & 261,359 & 320,540 & 215 & 215 & \textcolor{red}{207} & 215 & \textcolor{red}{207} & 189 & 189 & 199 & \textcolor{red}{181} & 191\\
\hline
cholesky\_bdti & 266,422 & 342,688 & \textcolor{red}{1156} & 1157 & \textcolor{red}{1156} & 1161 & \textcolor{red}{1156} & 947 & 1024 & 1120 & 1024 & \textcolor{red}{\underline{848}}\\
\hline
denoise & 275,638 & 356,848 & \textcolor{red}{416} & 722 & \textcolor{red}{416} & 516 & \textcolor{red}{416} & 224 & 478 & 244 & 224 & \textcolor{red}{\underline{220}}\\
\hline
sparcT2\_core & 300,109 & 302,663 & 1244 & 1273 & 1186 & 1319 & \textcolor{red}{\underline{1183}} & 1245 & 1972 & 1186 & 1081 & \textcolor{red}{\underline{918}}\\
\hline

gsm\_switch & 493,260 & 507,821 & 1827 & 5974 & 1759 & 1714 & \textcolor{red}{\underline{1621}} & \textcolor{red}{1407} & 5352 & 1719 & 1503 & \textcolor{red}{1407}\\
\hline
mes\_noc & 547,544 & 577,664 & \textcolor{red}{634} & 699 & 649 & 699 & 651 & \textcolor{red}{617} & 633 & 755 & 633 & \textcolor{red}{617}\\
\hline
LU230 & 574,372 & 669,477 & \textcolor{red}{3273} & 4070 & 4012 & 3452 & 3602 & \textcolor{red}{2677} & 3276 & 3751 & 2720 & 2923\\
\hline
LU\_Network & 635,456 & 726,999 & 525 & 550 & \textcolor{red}{524} & 550 & \textcolor{red}{524} & \textcolor{red}{524} & 528 & \textcolor{red}{524} & 528 & \textcolor{red}{524}\\
\hline
sparcT1\_chip2 & 820,886 & 821,274 & 899 & 1524 & 874 & 1129 & \textcolor{red}{\underline{873}} & 783 & 1029 & 856 & 877 & \textcolor{red}{\underline{757}}\\
\hline
directrf & 931,275 & 1,374,742 & \textcolor{red}{574} & 646 & 646 & 646 & 632 & \textcolor{red}{295} & 379 & \textcolor{red}{295} & 317 & \textcolor{red}{295}\\
\hline
bitcoin\_miner & 1,089,284 & 1,448,151 & \textcolor{red}{1297} & 1570 & 1576 & 1562 & 1514 & \textcolor{red}{1225} & 1255 & 1287 & 1255 & 1282\\
\hline
\multicolumn{3}{|c|}{Average Improvement over hMETIS ($\%$)} & 10.99 & 0 & 9.95 & 6.82 & \textcolor{red}{12.16} & 21.73 & 0 & 14.14 & 20.94 & \textcolor{red}{22.53}\\
\hline

\end{tabularx}
\end{table*}

\subsubsection{SHyPar Runtime Comparison with KaHypar}
Table {\ref{tab:runtime}} presents a runtime comparison between SHyPar and KaHyPar on the Titan23 benchmark for $\epsilon$ = $2\%$. The setup time corresponds to the one-time computation of the effective resistance (algorithm \ref{alg:effR}) and the flow-based clustering method (algorithm \ref{alg:flow_based}). The SHyPar/KaHyPar runtime ratio compares the runtime of SHyPar with the KaHyPar method, excluding the one-time setup time of effective resistance computation and flow-based clustering in SHyPar. To ensure a fair comparison, the runtime of Louvain clustering is also excluded from KaHyPar. The results demonstrate that our method achieves up to $8\%$ improvement in runtime over KaHyPar.

\begin{table}[ht]
% \tiny    
\caption{SHyPar runtime comparison with KaHyPar} \label{tab:runtime}
\centering
\footnotesize
\setlength{\tabcolsep}{4.5pt}
\begin{tabular}{|c|c|c|c|}
\hline
Benchmark &  \begin{tabular}{@{}c@{}} KaHyPar \\
(s)
\end{tabular}& \begin{tabular}{@{}c@{}} Setup Time \\
(s)
\end{tabular} & \(\displaystyle \frac{\text{SHyPar}}{\text{KaHyPar}}\) \\
\hline
sparcT1\_core & 52.99 & 10.19 & 0.94 \\
\hline
neuron & 41.06 & 10.71 & 0.93 \\
\hline
stereo\_vision & 3.19 & 10.63 & 0.99 \\
\hline
des90 & 49.90 & 13.29 & 0.94 \\
\hline
SLAM\_spheric & 10.12 & 12.37 & 0.99 \\
\hline
cholesky\_mc & 7.18 & 12.15 & 0.99 \\
\hline
segmemtation & 8.79 & 13.92 & 0.99\\
\hline
bitonic\_mesh & 43.52 & 18.04 & 0.93 \\
\hline
dart & 47.19 & 16.80 & 0.94 \\
\hline
openCV & 47.23 & 18.07 & 0.94 \\
\hline
stap\_qrd & 22.52 & 19.28 & 0.93 \\
\hline
minres & 23.65 & 20.25 & 0.93 \\
\hline
cholesky\_bdti & 35.40 & 20.86 & 0.92\\
\hline
denoise & 22.66 & 23.57 & 0.93 \\
\hline
sparcT2\_core & 49.21 & 22.05 & 0.94\\
\hline
gsm\_switch & 139.76 & 32.38 & 0.98\\
\hline
mes\_noc & 74.12 & 36.82 & 0.96 \\
\hline
LU230 & 1071.02 & 38.26 & 1.00 \\
\hline
LU\_Network & 78.99 & 43.66 & 0.96 \\
\hline
sparcT1\_chip2 & 167.14 & 50.63 & 0.98 \\
\hline
directrf & 80.89 & 68.07 & 0.96 \\
\hline
bitcoin\_miner & 471.46 & 71.27 & 0.99 \\
\hline
\end{tabular}
\end{table}

\subsubsection{Ablation Study}
To evaluate the impact of spectral coarsening and flow-based community detection, we consider a new configuration that employs only spectral coarsening within KaHyPar. Table {\ref{tab:Ablation}} presents the cut size results for some benchmarks for $\epsilon$ = $2\%$. While replacing KaHyPar’s default coarsening with spectral coarsening (SC) already outperforms the baseline, adding flow-based community detection (Flow-Based CD) on top yields even better solutions.

\begin{table}[ht]
% \tiny    
\caption{Cut size improvement with Spectral Coarsening} \label{tab:Ablation}
\centering
\footnotesize
\setlength{\tabcolsep}{3.7pt}
\begin{tabular}{|c|c|c|c|}
\hline
Benchmark & KaHyPar & KaHyPar with SC & \begin{tabular}{@{}c@{}} KaHyPar with \\ SC and Flow-Based CD
\end{tabular} \\
\hline
dart & 924 & 803 & 784\\
\hline
OpenCV & 560 & 549 & 499\\
\hline
sparcT2\_core & 1186 & 1184 & 1183\\
\hline
gsm\_switch  & 1759 & 1732 & 1621\\
\hline
LU230 & 4012 & 3610 & 3602\\
\hline
IBM05 & 1712 & 1709 & 1707\\
\hline
IBM07 & 894 & 892 & 882\\
\hline
IBM08 & 1157 & 1146 & 1140\\
\hline
IBM10 & 1318 & 1272 & 1254\\
\hline
IBM18 & 1915 & 1551 & 1521\\
\hline

\end{tabular}
\end{table}

\section{Conclusion}
In this study, we introduced SHyPar, a multilevel hypergraph partitioning framework that enhances partitioning solutions and surpasses earlier studies in performance. We developed an innovative algorithm that incorporates spectral hypergraph coarsening techniques, leverages hyperedge effective resistances and flow-based community detection. Our comprehensive experimental analysis, conducted on real-world VLSI test cases, demonstrates that SHyPar consistently achieves a significant reduction in hypergraph partitioning cut size, improving results up to 15 percent compared to state-of-the-art methods.

\section{Acknowledgments}
This work is supported in part by  the National Science Foundation under Grants    CCF-2417619, CCF-2021309,  CCF-2011412, CCF-2212370, and CCF-2205572.


\bibliographystyle{ieeetr}
\bibliography{SHyPar}

\begin{IEEEbiography}[{\includegraphics[width=1in,height=1.25in,clip,keepaspectratio]{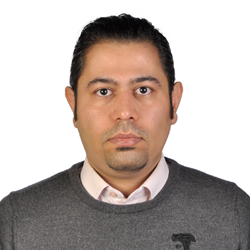}}]{Hamed Sajadinia}
received the M.Sc. degree in telecommunication engineering from Iran University of Science and Technology, Tehran, Iran, in 2016. He is currently pursuing the Ph.D. degree in Electrical and Computer Engineering with Stevens Institute of Technology, Hoboken, USA.

His research interests include VLSI Design, Machine Learning, and Graph-related problems.
\end{IEEEbiography}
\begin{IEEEbiography}[{\includegraphics[width=1in,height=1.25in,clip,keepaspectratio]{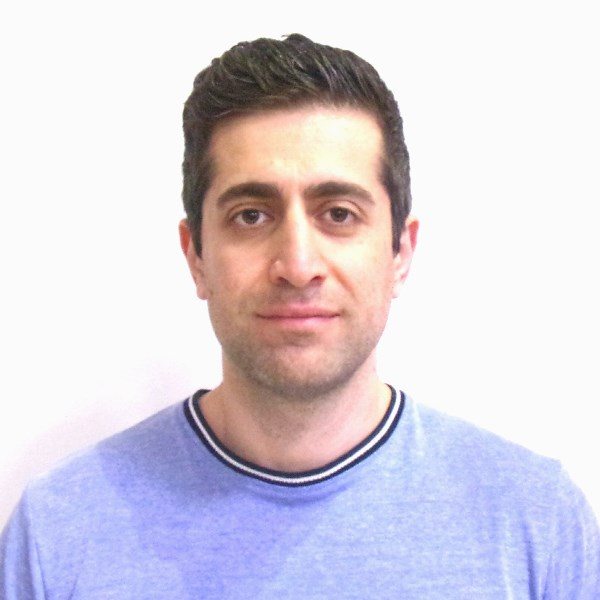}}]{Ali Aghdaei}
received the M.Sc. degree in electrical and computer engineering from Michigan Technological University, Houghton, MI, in 2016, and the Ph.D. degree in electrical and computer engineering from Stevens Institute of Technology, Hoboken, NJ, in 2023. He is currently a postdoctoral scholar at the University of California, San Diego, where he is a member of the ABKGroup research team. His research interests include electronic design automation (EDA), very large-scale integration (VLSI) design, and spectral (hyper)graph algorithms.

He received the Best Ph.D. Dissertation Award from the Department of Computer Engineering at Stevens Institute of Technology in 2023. He has also served as a Technical Program Committee (TPC) member for the ACM/IEEE Great Lakes Symposium on VLSI (GLSVLSI) 2025 conference.
\end{IEEEbiography}

\begin{IEEEbiography}[{\includegraphics[width=1in,height=1.25in,clip,keepaspectratio]{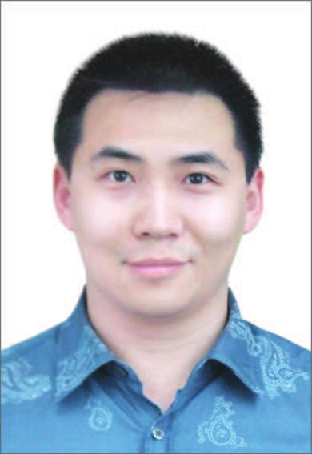}}]{Zhuo Feng} (S'03-M'10-SM'13) received the B.Eng. degree
in information engineering from Xi'an Jiaotong University, Xi'an,
China, in 2003, the M.Eng. degree in electrical engineering from
National University of Singapore, Singapore, in 2005, and the Ph.D.
degree in electrical and computer engineering from Texas A\&M
University, College Station, TX, in 2009. He is currently an associate professor at Stevens Institute of Technology. His
research interests include high-performance spectral methods, very large scale integration
(VLSI) and computer-aided design (CAD), scalable hardware and software systems,  as well as heterogeneous parallel computing.

He received a Faculty Early Career Development (CAREER)
Award from the National Science Foundation (NSF) in 2014, a Best
Paper Award from ACM/IEEE Design Automation Conference (DAC) in
2013, and two Best Paper Award Nominations from IEEE/ACM International
Conference on Computer-Aided Design (ICCAD) in 2006 and 2008. He was the principle investigator of the CUDA Research Center named by Nvidia Corporation. He has
served on the technical program committees of major international conferences
related to electronic design automation (EDA), including DAC, ASP-DAC,
ISQED, and VLSI-DAT, and has been a technical referee for many leading
IEEE/ACM journals in VLSI and parallel computing.  In 2016, he became a co-founder of LeapLinear Solutions to provide highly scalable software solutions for solving sparse matrices and analyzing graphs (networks) with billions of elements, based on the latest breakthroughs in spectral graph theory.
\end{IEEEbiography}

% \end{thebibliography}

\end{document}